\documentclass[journal]{IEEEtran}
  
\usepackage{xcolor} 
\usepackage{amsmath,amssymb,amsfonts,bbm} 
\usepackage{algorithmic}
\usepackage{tabularx} 
\usepackage{amsmath, bm}
\usepackage{amsmath,autobreak} 
\usepackage[ruled, linesnumbered]{algorithm2e} 
\usepackage{subfigure} 
\usepackage{soul} 
\usepackage{balance} 
\usepackage{color} 
\usepackage{float}
\usepackage{autobreak} 
\usepackage{graphicx} 
\usepackage{textcomp} 
\usepackage{xcolor} 
\usepackage{booktabs} 
\usepackage{multirow} 
\usepackage{bbm} 
\usepackage{dsfont} 
\usepackage{footnote} 
\usepackage[]{mathtools} 
\usepackage{lipsum} 
\makesavenoteenv{algorithm} 
\usepackage{diagbox}
\usepackage{lineno}
\usepackage{tikz}
\usepackage{makecell}
\usepackage{amssymb}
\usepackage{mathrsfs}
\usepackage{calligra}

\usepackage[numbers,sort&compress]{natbib}
\usepackage[colorlinks, linkcolor=black, anchorcolor=black, citecolor=black]{hyperref}

\newcommand*{\circled}[1]{\lower.7ex\hbox{\tikz\draw (0pt, 0pt)%
    circle (.5em) node {\makebox[1em][c]{\small #1}};}}

\allowdisplaybreaks[4]

\begin{document}
%
\title{
Hybrid-Task Meta-Learning: A GNN Approach for Scalable and Transferable Bandwidth Allocation
}

\author{    
Xin~Hao, 
Changyang~She,~\IEEEmembership{Senior Member,~IEEE,}
Phee~Lep~Yeoh,~\IEEEmembership{Senior Member,~IEEE},
Yuhong~Liu,
Branka~Vucetic,~\IEEEmembership{Life~Fellow,~IEEE,} 
and~Yonghui~Li,~\IEEEmembership{Fellow,~IEEE}.

\thanks{
The work of C. She was supported in part by DECRA under Grant DE210100415. The work of P. L. Yeoh was supported in part by ARC under Grant DP190100770. The work of Branka Vucetic was supported in part by the ARC Laureate Fellowship grant number FL160100032. The work of Yonghui Li was supported by ARC under Grant DP210103410.
This article was presented in part at the 2023 IEEE International Conference on Communications Workshops (ICC workshops)~\cite{Xin_gnn_conf}. \emph{(Corresponding author: Changyang She.)}
}


\thanks{

Changyang She is with the School of Electronic and Information Engineering, Harbin Institute of Technology (Shenzhen), Shenzhen 518055, China. Part of this work was done when he was with the School of Electrical and Information Engineering, The University of Sydney, Sydney, NSW 2006, Australia (e-mail: shechangyang@gmail.com).

P. L. Yeoh is with the School of Science, Technology and Engineering, University of the Sunshine Coast, QLD 4502, Australia (e-mail: pyeoh@usc.edu.au).

Y. Liu, B. Vucetic, and Y. Li are with the School of Electrical and Computer Engineering, University of Sydney, NSW 2006, Australia (e-mail: yuhong.liu@sydney.edu.au; branka.vucetic@sydney.edu.au; yonghui.li@sydney.edu.au).
} 


}

\markboth{
Accepted by IEEE Transactions on Wireless Communications
}%
{Shell \MakeLowercase{\textit{et al.}}: Bare Demo of IEEEtran.cls for IEEE Journals}   

\maketitle
\begin{abstract}
In this paper, we develop a deep learning-based bandwidth allocation policy that is: 1) scalable with the number of users and 2) transferable to different communication scenarios, such as non-stationary wireless channels, different quality-of-service (QoS) requirements, and dynamically available resources. To support scalability, the bandwidth allocation policy is represented by a graph neural network (GNN), with which the number of training parameters does not change with the number of users. To enable the generalization of the GNN, we develop a hybrid-task meta-learning (HML) algorithm that trains the initial parameters of the GNN with different communication scenarios during meta-training. Next, during meta-testing, a few samples are used to fine-tune the GNN with unseen communication scenarios. Simulation results demonstrate that our HML approach can improve the initial performance by $8.79\%$, and sample efficiency by $73\%$, compared with existing benchmarks. After fine-tuning, our near-optimal GNN-based policy can achieve close to the same reward with much lower inference complexity compared to the optimal policy obtained using iterative optimization. {Numerical results validate that our HML can reduce the computation time by approximately $200$ to $2000$ times than the optimal iterative algorithm.}



\end{abstract}
\begin{IEEEkeywords}
Bandwidth allocation, graph neural network, meta-learning, quality-of-service.
\end{IEEEkeywords}

\IEEEpeerreviewmaketitle

\section{Introduction}
\IEEEPARstart{T}{hroughout} the rapid evolution of wireless communication systems, the spectral efficiency, which is the amount of information that can be transmitted over a given bandwidth while maintaining a certain quality of service (QoS) level, still remains one of the most critical performance metrics for future sixth-generation (6G) wireless communications~\cite{Xin_gnn_conf, 6G_ComSoc}. To maximize spectrum efficiency, low-complexity bandwidth allocation solutions are critical for real-time decision-making within each transmission time interval (TTI) that could be shorter than one millisecond in current fifth-generation (5G) wireless communications. Furthermore, the number of users requesting bandwidth in each TTI is stochastic~\cite{GYF_GNN_TWC, TWC_GNN_YCY}, each user may have different QoS requirements~\cite{TWC_NS_multiservice, TWC_Heterogeneous_service, TWC_LACO}, and wireless channels are non-stationary~\cite{TWC_transfer_meta, fast_meta_TWC}, making it difficult to develop a low-complexity bandwidth allocation policy that is scalable with the number of users and can satisfy a diverse range of communication scenarios.


Existing iterative optimization algorithms can obtain optimal bandwidth allocation policies, but their computational complexity is generally too high to be implemented in real time~\cite{SCY_transfer, TWC_BA_MP_iter, TWC_BA_iter_SGD}. To reduce the computational complexity, deep learning is a promising approach for 6G communications~\cite{SCY_tutorial_urllc, Xin_BCDRL_TCOM}. The idea is to train a deep neural network that maps the network status to the optimal decision. After training, the deep neural network can be used in communication systems for real-time decision-making, referred to as inference~\cite{Quek_SR_DRL_noise}. Although deep learning has much lower inference complexity compared with iterative optimization algorithms, existing deep learning solutions using fully connected neural networks (FNNs) are not scalable to different number of users in wireless networks~\cite{SCJ_Book_Chapter}. This is because the number of training parameters of an FNN depends on the dimensions of the input and output, which change with the number of users. Thus, a well-trained FNN is not applicable in wireless networks with stochastic user requests. In contrast to FNNs, graph neural networks (GNNs) have scalable numbers of training parameters that adapt to the number of users~\cite{MPNN_chemistry} --- making them highly-suitable for developing scalable deep learning-based resource allocation solutions for wireless networks~\cite{MPGNN_HK_JSAC, LYH_GNN_interference}. Furthermore, improving the generalization ability of GNN in wireless networks with diverse QoS requirements remains an open problem.

{When the services with diverse QoS requirements co-exist in a network, the first step is to reserve physical resources among different types of services by network slicing (NS) technology for each slice~\cite{Xin_ICC2024}, aiming to support}
diverse QoS requirements, such as data rate~\cite{TWC_max_sum_rate_noAI, TCOM_data_rate_general_DNN_YCY}, latency~\cite{TWC_EC_2003, TWC_EC_2007}, and security~\cite{Niyato_SR_DRL_IRS_TWC, TWC_joint_Pout_Rs_Quek, TWC_SR_Poor, TWC_Rs_short_Poor}, in both long and short coding blocklength regimes~\cite{TWC_NS_DRL_eMBB_URLLC, SCY_Rs_short, TIT_short_Poor}. 
To reserve resources for a single slice, the authors of~\cite{Access_NS_QCI} proposed to compute the weights of different slices based on the corresponding QoS requirements and the number of service requests. With this approach, the amount of reserved resources for each slice is stochastic. Meanwhile, since the wireless channels are non-stationary, the reserved resources and the wireless channels in the training stage could be different from the actual required resources in the testing stage~\cite{TWC_channel_mismatch_CSI, JSAC_power_allocation_mismatch}. As such, the mismatch between training data samples and testing data samples remains a crucial bottleneck for implementing efficient learning-based policies in practical wireless networks.


{Recent works have considered to reduce the online training time using transfer learning, which involves offline pre-training of the neural network in a known scenario and online fine-tuning to apply it to a related scenario~\cite{SCY_transfer}. 
Compared with conventional random initialization, transfer learning can enhance sample efficiency in a target wireless communication scenario by leveraging well-trained neural network features from a related communication scenario. However, when considering a large number of communication scenarios including the inherent non-stationary nature of wireless communications, transfer learning will require more time to train the neural network due to limited data samples in each communication scenario.
To further improve the online training efficiency for unseen wireless communication scenarios}, meta-learning has been proposed~\cite{bySGD_bySGD, MAML_Finn, Reptile, Feature_Reuse}. One of the meta-learning algorithms, model-agnostic meta-learning (MAML), has been applied to solve policy mismatch issues caused by varying user requests and non-stationary wireless channels~\cite{meta_learn_THzVR_TWC, MetaMEC_letter, TWC_transfer_meta, fast_meta_TWC}. While these aforementioned works have highlighted the generalization ability of meta-learning for non-stationary wireless resource allocation, no works have addressed the impact of diverse QoS requirements in {wireless} communications.

In this paper, we put forth a low-complexity bandwidth allocation framework by designing a GNN that is scalable with the number of users and applying meta-learning to generalize the GNN to different communication scenarios.
The main contributions are summarized as follows,
\begin{itemize}
\item Our proposed GNN is designed to handle six diverse QoS requirements of data rate, latency, and security in each of the long and short coding blocklength regimes. This generalization is achieved by using feature engineering to translate the channel state information (CSI) and customized QoS requirement of individual users into the minimum required bandwidth. 
\item Based on the extracted feature of minimum required bandwidth, we design a GNN-based bandwidth allocation policy that is scalable to the number of users. To train the GNN, we apply an unsupervised learning method to maximize the sum reward of the users with different QoS requirements in a network-slicing architecture. 
\item The optimal bandwidth allocation policies are obtained based on an iterative optimization algorithm to obtain the performance limit of the GNN-based policy in terms of the sum reward. By analyzing the computational complexity, we show that the GNN has a much lower inference complexity compared with the iterative optimization algorithm that is optimal.
%
\item Finally, we develop our generalized hybrid-task meta-learning (HML) algorithm that is transferable to different communication scenarios by using meta-training to train the initial parameters of the GNN. We note that only a few samples are required to fine-tune the parameters of the GNN in meta-testing which validates that our GNN-based policy initialized by HML can be efficiently transferred to previously unseen communication scenarios. Simulation results show that our GNN-based policy achieves near-optimal performance and HML significantly outperforms the three considered benchmarks of MAML, MTL transfer (multi-task learning based transfer learning), and random initialization. 
\end{itemize}

In our simulations,
the sum reward achieved by the GNN-based policy and that of the optimal bandwidth allocation policy obtained from the iterative optimization algorithm {can be overlapped in the long blocklength regime. The performance gap in between is less than $6\%$ even in the scenario considering effective capacity that the iterative algorithm has considerably higher computational complexity.} HML also improves the initial performance by up to $8.79\%$ and sample efficiency by up to $73\%$ compared with the MAML benchmark. We also show that the performance gains of HML is even higher when compared to the other two benchmarks.

\section{Related Works}
\subsection{Deep Learning for Resource Allocation in Wireless {Networks}}
Applying deep learning for resource allocation in wireless networks has been widely studied in the existing literature~\cite{Quek_SR_DRL_noise, SCJ_Book_Chapter}. In~\cite{Quek_SR_DRL_noise}, the authors showed that learning-based algorithms could obtain near-optimal solutions {by using a deep FNN}, and the computational complexity in inference is low. In~\cite{SCJ_Book_Chapter}, the authors proposed an FNN-based unsupervised learning algorithm to optimize the bandwidth allocation policy. More recently, due to the fact that FNN is not scalable to the number of users, GNNs have been applied in wireless networks optimizations~\cite{MPGNN_HK_JSAC, LYH_GNN_interference}. In~\cite{LYH_GNN_interference}, the authors designed a GNN, which is scalable to the number of users in a wireless network, to minimize the summation of queuing delay violation probability and packet loss probability. In~\cite{MPGNN_HK_JSAC}, the authors developed GNN-based scalable learning-based methods to solve radio resource management problems. {How to efficiently generalize a scalable GNN to allocate bandwidth resources across different communication scenarios remains an open issue.}

\begin{table}[t] 
\renewcommand\arraystretch{0.7} 
\caption{Considered QoS Requirements in Related Works} 
\centering 
\begin{tabular}{l | c c | c c | c c}
\toprule 
\toprule 
\multirow{2}{*}{\diagbox{\textbf{Refs}}{\textbf{QoS}}} & \multicolumn{2}{c|}{\makecell{\textbf{Data rate}}}                      & \multicolumn{2}{c|}{\makecell{\textbf{Latency}}} & \multicolumn{2}{c}{\makecell{\textbf{Security}}}\\
\cline{2-7}\rule{0pt}{9pt} 
                                                        & \makecell{Long} & \makecell{Short} & \makecell{Long}      & \makecell{Short} & \makecell{Long} & \makecell{Short} \\
\midrule 
\cite{TWC_max_sum_rate_noAI,TCOM_data_rate_general_DNN_YCY}                            &$\checkmark$ & & & & & \\
\hline
\cite{TWC_EC_2003, TWC_EC_2007}                         & & &$\checkmark$ & & & \\
\hline
\cite{TWC_SR_Poor, Niyato_SR_DRL_IRS_TWC, TWC_joint_Pout_Rs_Quek}    & & & & &$\checkmark$ & \\
\hline
\cite{TWC_Rs_short_Poor}                                & & & & & &$\checkmark$ \\
\hline
\cite{SCY_Rs_short}                                     & & & &$\checkmark$ & &$\checkmark$ \\
\hline
\cite{TWC_NS_DRL_eMBB_URLLC}                            &$\checkmark$ &$\checkmark$ & & & & \\
\hline
\cite{SCY_transfer}                                     &$\checkmark$ &$\checkmark$ &$\checkmark$ & & & \\
\bottomrule 
\bottomrule 
\end{tabular} 
\label{table_literature_review} 
\end{table}




\subsection{Generalization of Deep Learning 
Policies 
in Non-Stationary Wireless Networks
}
In wireless networks, the user requests, wireless channels, and available resources for each type of service can be non-stationary. Table~\ref{table_literature_review} summarizes some QoS requirements considered in the related works. For example, data rate, latency, and security have been investigated in~\cite{TCOM_data_rate_general_DNN_YCY, TWC_max_sum_rate_noAI, TWC_EC_2003, TWC_EC_2007, TWC_SR_Poor, Niyato_SR_DRL_IRS_TWC, TWC_joint_Pout_Rs_Quek}. These papers mainly focus on scenarios with long channel coding blocklengths, where the achievable rate of a wireless link can be approximated by the Shannon capacity. In 5G {communications}, the coding blocklength can be short, and Shannon capacity is not applicable. As such, the authors of~\cite{TWC_Rs_short_Poor, SCY_Rs_short} established to optimize wireless communication systems using the achievable rate in the short blocklength regime~\cite{TIT_short_Poor}. {In addition, since} different services may co-exist in one network, the authors of~\cite{TWC_NS_DRL_eMBB_URLLC, SCY_transfer} considered different QoS requirements in both long and short blocklength regimes. To support diverse QoS requirements in network slicing, the authors of~\cite{Access_NS_QCI} proposed to reserve bandwidth for different slices based on the number of users and the required QoS. {As shown in Table~\ref{table_literature_review}, none of the previously mentioned works proposed a general solution that can simultaneously support data rate, latency, and security in both long and short blocklength regimes.}

Considering that the number of requests, the reserved resources, and the wireless channels {all can be} dynamic, improving the generalization ability of deep learning policies has attracted significant research interests in recent years. One approach to address this challenge is to {deliberately} initialize the neural network and fine-tune it online. The authors of~\cite{SCY_transfer} applied transfer learning to fine-tune the parameters of deep neural networks that are trained offline in dynamic wireless networks, {thus to improve the sample efficiency when the communication scenario changed to the other one}. To further improve the sample efficiency {across different unseen communication scenarios}, meta-learning has been adopted in~\cite{TWC_transfer_meta, fast_meta_TWC, MetaMEC_letter, meta_learn_THzVR_TWC}, where the hyper-parameters of a deep neural network, such as the initial parameters, are updated according to a set of communication scenarios in meta-training. In~\cite{MetaMEC_letter}, meta-learning was applied to optimize computing resource allocation policies in mobile edge computing networks to fit both time-varying wireless channels and different requests of computing tasks. In~\cite{meta_learn_THzVR_TWC}, meta-learning was applied in virtual reality to quickly adapt to the user movement patterns changing over time. To improve the training efficiency in non-stationary vehicle networks, the authors in~\cite{TWC_transfer_meta} proposed optimizing the beamforming using meta-learning. In~\cite{fast_meta_TWC}, the authors combined meta-learning and support vector regression to extract the features for beamforming optimization, further improving training efficiency over non-stationary channels. 
{There is no existing literature for using GNN-based meta-learning to implement network slices that support diverse communication scenarios.}

\section{System Model and Problem Formulation}
We consider an uplink orthogonal-frequency-division-multiple-access communication system with network slicing where $U$ users are requesting different types of services from one base station (BS). The BS first reserves bandwidth for each type of service according to the QoS requirement and the number of users. Then, it allocates bandwidth to different users within each slice. The resource reservation for different slices has been extensively studied in the existing literature, so we will focus on developing bandwidth allocation policies for individual slices with different numbers of users, non-stationary wireless channels, and dynamic available bandwidth.



\subsection{Bandwidth Reservation for {Individual} Slices}
We assume that there can be multiple bandwidth reservation policies for different slices in network slicing (NS), and users of the same QoS requirement and {the same} blocklength share the same slice. We further use the QoS class identifier (QCI) to characterize the users in the slices~\cite{Access_NS_QCI}. Given the total bandwidth of the BS, $W_{\max}$, the bandwidth reserved for the $\tau$-th slice is given by
\begin{equation}
    W_{\tau}^{\Phi,\xi} = f_{\tau}^\mathrm{NS}\left(U_{\tau}^{\Phi,\xi}, I_{u_{\tau}}^{\Phi,\xi}\right) \cdot W_{\max},
    \label{eq_reserved_bandwidth}
\end{equation}
{where superscript $\Phi$ represents the QoS requirement in the current network slice, whilst the superscript $\xi$ represents whether the current scenario is in a long or short blocklength regime, respectively. Function $f_{\tau}^\mathrm{NS}\left(\cdot,\cdot\right)$ is the NS} function for bandwidth reservation in NS, $U_{\tau}^{\Phi,\xi}$ is the number of users in the $\tau$-th slice {with QoS requirement as $\Phi$ and in long or short blocklength regime represented by $\xi$, and} $I_{u_{\tau}}^{\Phi,\xi}$ is the QCI of the $u_{\tau}$-th user in the $\tau$-th slice. Since the sum of the bandwidth reserved for all the slices equals the total bandwidth of the BS, thus {can be described as}
\begin{equation}
    \sum_{\tau=1}^{T_\mathrm{NS}} f_{\tau}^\mathrm{NS}\left(U_{\tau}^{\Phi,\xi},I_{u_{\tau}}^{\Phi,\xi}\right) = 1.
\end{equation}
where $T_\mathrm{NS}$ is the number of {total network slices managed by the BS}. Inspired by~\cite{Access_NS_QCI}, the bandwidth reserved for each slice depends on the number of users in this slice and the QCI of these users, e.g., 
\begin{equation}
    f_{\tau}^\mathrm{NS}(\cdot) = \frac{\sum_{u\in \mathcal{U}_{\tau}^{\Phi,\xi}} I_{u_{\tau}}^{\Phi,\xi}}{\sum_{\tau=1}^{T_\mathrm{NS}} \sum_{u \in \mathcal{U}_{\tau}^{\Phi,\xi}} I_{u_{\tau}}^{\Phi,\xi}}.
\end{equation}

\subsection{Different QoS in {Long} and {S}hort {B}locklength Regimes}
To investigate the generalization ability of our proposed bandwidth allocation policy, we consider both long and short blocklength regimes with three types of QoS requirements, i.e., data rate, queuing delay, and security. Thus, there are six communication objectives in total. {We present details of these six objectives from the following three aspects.}

\subsubsection{Data Rate Requirement}
When the blocklength is long, the data rate reward of the $u$-th user can be expressed as~{\cite{SCY_transfer}}
\begin{equation}
r_u^{D,\mathcal{I}} = \frac{w_u}{\ln 2} \ln\left(1 + \frac{P_u h_u}{w_u N_0}\right),
\label{eq_user_data_rate}
\end{equation}
where $w_u$ is the bandwidth allocated to the $u$-th user, $P_u$ is the transmit power of the $u$-th user, $N_0$ is the single-sided noise spectral density, and $h_u=\alpha_u^{} g_u^{}$ is the channel gain, where  ${\alpha}_u^{}$ and $g_u^{}$ represent the large-scale and small-scale channel gains between the $u$-th user and the BS, respectively.

When the blocklength is short, decoding errors cannot be neglected. As such, the data rate reward of the $u$-th user can be approximated by~\cite{TIT_short_Poor}
\begin{equation}
    r_u^{D,\mathcal{F}} 
    \approx r_u^{D,\mathcal{I}} -\sqrt{\frac{V_u}{L_u}} \frac{f_{Q}^{-1}(\epsilon_{u}) }{\ln2/w_u}
    \label{eq_data_rate_short}
\end{equation}
where $V_u = 1-{\left(1+\frac{P_u h_u}{w_u N_0}\right)^{-2}}$ is the channel dispersion that measures the stochastic variability of the channel related to a deterministic channel with the same capacity, $L_u=T_\mathrm{s} w_u$ is the blocklength, and $T_\mathrm{s}$ is the transmission duration of each coding block. The function $f_{Q}^{-1}(x)$ is the inverse of the Gaussian Q-function, and $\epsilon_{u}$ is the decoding error probability.

\subsubsection{Latency Requirement}
When considering latency constraints due to queueing delays, the effective capacity is applied to characterize the statistical QoS requirement in wireless communications. {The effective capacities in long and short blocklength regimes are expressed as}
{
~\cite{SCY_Rs_short}
\begin{equation}
    r_u^{E,\mathcal{I}} 
   = -\frac{1}{\vartheta_u T_\mathrm{c}} \ln\left(\mathbb{E}_{g_u}\left[\exp\left(-\vartheta_u T_\mathrm{c} r_u^{D,\mathcal{I}} \right)\right]\right),
   \label{eq_user_effective_rate_long}
\end{equation}
and
\begin{equation}
    r_u^{E,\mathcal{F}} 
   = -\frac{1}{\vartheta_u T_\mathrm{c}} \ln\left(\mathbb{E}_{g_u}\left[\exp\left(-\vartheta_u T_\mathrm{c} r_u^{D,\mathcal{F}} \right)\right]\right),
   \label{eq_user_effective_rate_short}
\end{equation}
respectively,
}
where $T_\mathrm{c}$ is the channel coherence time, $\vartheta_u$ is the QoS exponent for queuing delay, $\mathbb{E}[\cdot]$ denotes the expectation, and $r_u^{D,\xi}$ is the data rate in~\eqref{eq_user_data_rate} or~\eqref{eq_data_rate_short}. 
We note that $\vartheta_u = \frac{\ln(1/\varepsilon_u)}{a_u  {\tau}_{\max}}$ is determined by the maximum tolerable delay bound violation probability, $\varepsilon_u$, the packet arrival rate, $a_u$, and the threshold of queuing delay, $\tau_{\max}$. 

\subsubsection{Security Requirement}
To formulate the wireless security requirement, we consider that there is an eavesdropper that attempts to {eavesdrop} the information transmitted by each user. 

In the long blocklength regime, the secrecy rate of the $u$-th user can be expressed as~\cite{TWC_SR_Poor}
\begin{equation}
r_u^{S,\mathcal{I}} = \left[r_u^{D,\mathcal{I}} - r_u^{e,\mathcal{I}} \right]^{+}, 
\label{eq_secrecy_rate}
\end{equation}
where $[x]^{+}=\max\{0,x\}$, and $r_u^{e,\mathcal{I}} = \frac{w_u}{\ln 2}\ln\left(1+ \frac{P_u h_u^{e}} {w_u N_0}\right)$ is the data rate of the {eavesdrop}ped channel from the $u$-th user to the eavesdropper. The channel gain of the {eavesdrop}ped channel is denoted by $h_u^{e}=\alpha_u^{e} g_u^{e}$, where ${\alpha}_u^{e}$ and $g_u^{e}$ represent the large-scale and small-scale channel gains between the $u$-th user and the eavesdropper, respectively. 

In the short blocklength regime, the achievable secrecy rate of the $u$-th user can be approximated as~\cite{TWC_Rs_short_Poor},
\begin{align}
r_u^{S, \mathcal{F}} = 
\begin{cases} 
r_u^{S,\mathcal{I}} 
   - \sqrt{\frac{V_u}{L_u}} \frac{f_{Q}^{-1}(\epsilon_{u})}{\ln 2/w_u}
   - \sqrt{\frac{V_u^{e}}{L_u}} \frac{f_{Q}^{-1}(\delta_{u})}{\ln 2/w_u},     &h_u >     h_u^{e}\\ 
0,                                                                        &h_u \leq  h_u^{e},
\end{cases} 
\label{eq_secrecy_rate_short} 
\end{align} 
where $V_u^{e} = 1-{\left(1+\frac{P_u h_u^{e}}{w_u N0}\right)^{-2}}$, and $\delta_{u}$ represents the information leakage, which describes the statistical independence between the transmitted confidential message and the eavesdropper's observation, and is measured by the total variation distance~\cite{TWC_Rs_short_Poor}. 

\subsection{Problem Formulation}
To maximize the sum reward subject to the QoS requirements {and blocklength} in each slice, we formulate the bandwidth allocation problem as follows,
\begin{align}
\max_{\boldsymbol{w}_{\tau}^{\Phi,\xi}} \quad & \sum_{u \in \mathcal{U}_{\tau}^{\Phi,\xi}} r_u^{\Phi, \xi},                                            \label{eq_problem_variable} \\
\mathrm{s.t.} \quad \nonumber
& \sum\limits_{u \in \mathcal{U}_{\tau}^{\Phi,\xi}} w_u^{\Phi,\xi} \leq W_{\tau}^{\Phi,\xi},                 \tag{\ref{eq_problem_variable}{a}} \label{eq_problem_variable_a}
\\
& w_{u}^{\Phi,\xi} \geq 0 ,                                                        \tag{\ref{eq_problem_variable}{b}} \label{eq_problem_variable_b}
\\
& r_u^{\Phi,\xi} \geq r_{\tau}^{\Phi,\xi},                                      \tag{\ref{eq_problem_variable}{c}} \label{eq_problem_variable_c}
\end{align}
where $\boldsymbol{w}_{\tau}^{\Phi,\xi}=[w_1^{\Phi,\xi}, w_2^{\Phi,\xi}, \cdots, w_{U_{\tau}}^{\Phi,\xi}]^\mathrm{T}$ is the bandwidth allocated to the users with QoS requirement and $r_{\tau}^{\Phi,\xi}$ is the minimum threshold of the QoS required by the users, {where superscripts $\Phi$ represents the QoS requirement in the current network slice, and $D, E, S$ indicate data rate, effective capacity with queuing delay constraint, and secrecy rate, respectively; whilst the superscripts {$\xi$ represents whether the current scenario is in a long or short blocklength regime, and $\mathcal{I, F}$ indicate} the scenarios in the infinite long and finite short blocklength regimes, respectively}. Thus, constraint~\eqref{eq_problem_variable_c} guarantees the QoS of all the users. Taking eqs.~\eqref{eq_user_data_rate}--\eqref{eq_secrecy_rate_short} into problem~\eqref{eq_problem_variable_a}, we can observe that problem~\eqref{eq_problem_variable} is non-convex, since four out of the six communication scenarios are non-convex.

\begin{algorithm}[t] 
\algsetup{linenosize=\normalsize} \small  
\caption{{User Scheduling Algorithm}}~\label{Algorithm_schedule_user}
{
Initialize the set of scheduled users as all the users: $\mathcal{K}_{\tau}^{\Phi,\xi}=\mathcal{U}_{\tau}^{\Phi,\xi}$.\\
\textit{1) Drop $u$-th user if the target in eq.~\eqref{eq_problem_variable_c} cannot be satisfied with $W_{\tau}^{\Phi,\xi}$}:\label{line_2_alg_1}\\
\For{$u \in \mathcal{U}_{\tau}^{\Phi,\xi}$}
{
Calculate the reward of each user by taking $w_u=W_{\tau}^{\Phi,\xi}$ into the target reward shown in eqs.~\eqref{eq_user_data_rate}--\eqref{eq_secrecy_rate_short} as:\label{line_reward_alg_1}
$r_{u,\max}^{\phi,\xi} = r_u^{\phi,\xi} (W_{\tau}^{\Phi,\xi}) $.\\
\uIf{$r_{u,\max}^{\phi,\xi} < r_{\tau}^{\Phi,\xi}$}
{
Drop the $u$-th user: $\mathcal{K}_{\tau}^{\Phi,\xi} = \mathcal{K}_{\tau}^{\Phi,\xi} - \{u\}$, $K = K-1$, and $w_u = 0$.\\
}
\Else
{Get $w_{k,\min}$ using binary search function:
$w_{k,\min} = f_\mathrm{BiS}(h_k, r_{\tau}^{\Phi,\xi})$. \label{algorithm_user_scheduling_line_w_k_min}
}
}
\textit{2) Drop $k$-th user if eq.~\eqref{eq_problem_variable_a} cannot be satisfied with $w_{k,\min}$}:\\
\While{$\sum\nolimits_{k =1}^{K} {w_{k,\min}} > W_{\tau}^{\Phi,\xi}  $~\label{algorithm_user_scheduling_line_constraint_c_begin}}
{
Identify user holding highest $w_{k,\min}$ in $\mathcal{K}_{\tau}^{\Phi,\xi}$:
$k_{\mathrm{drop}} = \arg\max\limits_{k} {w_{k, \min}}$. ~\label{algorithm_user_scheduling_line_argmax}\\
Drop the $k_\mathrm{drop}$-th user: $\mathcal{K}_{\tau}^{\Phi,\xi} = \mathcal{K}_{\tau}^{\Phi,\xi} - \{k_{\mathrm{drop}}\}$, $K = K-1$, and $w_{k_{\mathrm{drop}}}=0$.~\label{algorithm_user_scheduling_line_constraint_c_end}\\
}
}
\end{algorithm}

\subsection{{Analysis of Problem} Feasibility}
Given the available bandwidth constraint in~\eqref{eq_problem_variable_a} and the QoS constraint in~\eqref{eq_problem_variable_c}, problem~\eqref{eq_problem_variable} will be infeasible when some of the users in this slice have weak channels. We denote the minimum bandwidth required to meet constraint~\eqref{eq_problem_variable_c} by ${\boldsymbol{w}}_{\min}^{\Phi,\xi} = \left[{w}_{1,\min}^{\Phi,\xi}, \cdots, {w}_{U_{\tau},\min}^{\Phi,\xi}\right]^\mathrm{T}$. 
To get the minimum required bandwidth, we first assess if each user's reward meets constraint~\eqref{eq_problem_variable_c} by line~\ref{line_reward_alg_1} of Algorithm~\ref{Algorithm_schedule_user}. If not, it indicates this user is experiencing a deep fading channel, necessitating infinite bandwidth. In contrast, for users not experiencing deep fading, we employ a binary search function to compute their required minimum bandwidth. The step-by-step process is detailed in lines~\ref{line_2_alg_1}--10 of Algorithm~\ref{Algorithm_schedule_user}.
However, it is possible that the sum minimum bandwidth achieved in line~8 of Algorithm~1 is larger than $W_{\tau}^{\Phi,\xi}$, which is conflicted with constraint~\eqref{eq_problem_variable_a} conflicting with constraint~\eqref{eq_problem_variable_a}. Therefore, we must remove the users sequentially consuming the most bandwidth until constraint~\eqref{eq_problem_variable_a} is satisfied.
We note that if some users experience deep fading, leading to $\sum_{u \in \mathcal{U}_{\tau}^{\Phi,\xi}}{w}_{u,\min}^{\Phi,\xi} > W_{\tau}^{\Phi,\xi}$, then problem~\eqref{eq_problem_variable} is infeasible. In this case, the BS will only schedule the users with sufficiently strong channels. Alternatively, to maximize the number of scheduled users in problem~\eqref{eq_problem_variable}, we consider that the BS schedules the $K$ users with the {minimum} bandwidth requirement. Denote the set of scheduled users by $\mathcal{K}_{\tau}^{\Phi,\xi}$. Then, for any $k\in \mathcal{K}_{\tau}^{\Phi,\xi}$ and $u \notin\mathcal{K}_{\tau}^{\Phi,\xi}$, we have ${w}_{k,\min}^{\Phi,\xi} \leq {w}_{u,\min}^{\Phi,\xi}$.
{To achieve the scheduled users, we need to execute a user scheduling algorithm, and the step-by-step process is given in Algorithm~\ref{Algorithm_schedule_user}. With Algorithm~\ref{Algorithm_schedule_user}, we can optimize the number of scheduled users with the changes in the total number of requesting users. However, it's also practical to consider alternative user scheduling algorithms for different requirements, such as prioritizing users based on their historical usage patterns, implementing a round-robin scheduling approach to ensure equal distribution of resources among users, or incorporating machine learning techniques to adjust schedulings based on real-time traffic and user behavior dynamically.}

After user scheduling, problem~\eqref{eq_problem_variable} can be reformulated as follows,
\begin{align}
    \max_{\boldsymbol{w}_{\tau}^{\Phi,\xi}} \quad & \sum_{k \in \mathcal{K}_{\tau}^{\Phi,\xi}} {r_k^{\Phi,\xi}},                                        \label{eq_problem_variable_sche} \\
    \mathrm{s.t.} \quad \nonumber
    & \sum\limits_{k \in \mathcal{K}_{\tau}^{\Phi,\xi}} {w_k^{\Phi,\xi}} \leq W_{\tau}^{\Phi,\xi} ,      \tag{\ref{eq_problem_variable_sche}{a}} \label{eq_problem_variable_sche_a}\\
    & w_k^{\Phi,\xi} \geq w_{k,\min}^{\Phi, \xi}.                           \tag{\ref{eq_problem_variable_sche}{b}} \label{eq_problem_variable_sche_b} 
\end{align}
{We denote $K \leq U$ as the number of scheduled users.}



In the following, we investigate how to find the optimal solution to problem~\eqref{eq_problem_variable_sche}. 


\section{Hybrid-Task Meta-Learning for GNN-based Scalable Bandwidth Allocation}
In this section, we first illustrate how to obtain the optimal bandwidth allocation by using an iterative optimization algorithm. Next, we utilize feature engineering techniques to reformulate the problem, and represent the bandwidth allocation policy by a GNN. To generalize the GNN, the feature of {the} required minimum bandwidth that can be used to represent different QoS requirements is used as the GNN's input. Then, we develop a meta-learning approach to train the GNN. The goal is to obtain a policy that is scalable to the number of users and can generalize well in diverse communication scenarios with different channel distributions, QoS requirements, and available bandwidth.

\begin{algorithm}[t]
\algsetup{linenosize=\normalsize} \small  
\caption{Iterative Bandwidth Allocation Algorithm}\label{algorithm_bandwidth_allocation_incremental_search}
\textbf{Initialize}: Bandwidth of a resource block: {$\Delta w$}. \\
Use user scheduling algorithm to get the minimum required bandwidth for each scheduled user: $w_k^{\Phi,\xi} = w_{k,\min}^{\Phi,\xi}, \forall k \in \mathcal{K}_{\tau}^{\Phi,\xi}$.\\
\While{$ W_{\tau}^{\Phi,\xi} - \sum\nolimits_{k \in \mathcal{K}_{\tau}^{\Phi,\xi} } {w_k^{\Phi,\xi}} \geq \Delta w$}
{
\For{$k \in \mathcal{K}_{\tau}^{\Phi,\xi}$}
{
$\Delta r_k^{\Phi,\xi} (w_k^{\Phi,\xi})= r_k^{\Phi,\xi}(w_k^{\Phi,\xi} + \Delta w) - r_k^{\Phi,\xi} (w_k^{\Phi,\xi})$.~\label{iterative_DeltaW}
}
Identify user has highest $\Delta r_k^{\Phi,\xi}(w_k^{\Phi,\xi})$ in $\mathcal{K}_{\tau}^{\Phi,\xi}$: $k_{\mathrm{allo}} = \arg\max\limits_{k} {\Delta r_k^{\Phi,\xi} }(w_k^{\Phi,\xi})$.\\
Allocate extra $\Delta w$ bandwidth to the $k_\mathrm{allo}$-th user: $w_{k_\mathrm{allo}}^{\Phi,\xi} = w_{k_\mathrm{allo}}^{\Phi,\xi} + \Delta w$.
}
\textbf{Output}: Optimal bandwidth allocation policy: $\boldsymbol{w}^{\Phi,\xi, \mathrm{opt}}=\boldsymbol{w}^{\Phi,\xi}$.
\end{algorithm}

\subsection{Optimal Bandwidth Allocation by Iterative Algorithm}~\label{section_iterative_algorithm}
Inspired by the optimization algorithm for resource allocation in~\cite{SCY_transfer}, we propose an iterative optimization algorithm for solving our problems. We denote the bandwidth of each resource block by $\Delta w$. At the beginning of the iteration, the bandwidth allocated to each user is $w_{k,\min}^{\Phi, \xi}$. In each iteration, we calculate the incremental reward of each user when an extra resource block is allocated to it, denoted by $\Delta {r}_k^{\Phi,\xi} (w_k)= {r}_k^{\Phi,\xi}({w}_k^{\Phi,\xi} + \Delta {w}) - {r}_k^{\Phi,\xi} ({w}_k^{\Phi,\xi})$. Finally, the resource block is allocated to the user with the highest $\Delta {r}_k^{\Phi,\xi}$. The details of the algorithm can be found in Algorithm~\ref{algorithm_bandwidth_allocation_incremental_search}. 

The optimality of the algorithm depends on the properties of the problems. For problem~\eqref{eq_problem_variable_sche}, if it is a convex problem, then Algorithm~\ref{algorithm_bandwidth_allocation_incremental_search} can obtain the optimal solution~\cite{SCY_transfer}. To validate whether problem \eqref{eq_problem_variable_sche} is convex or not, we only need to validate whether $r_u^{\Phi, \xi}$ is concave or not. In the long blocklength regime, we can prove that the secrecy rate is concave in bandwidth. See proof in Appendix~\ref{appendix_concave}. Since Shannon's capacity is a special case of the secrecy rate when the {eavesdrop}ped channel is in deep fading, thus Shannon's capacity is also concave in bandwidth. In addition, the authors of~\cite{TWC_EC_concave} proved that the effective capacity is concave in bandwidth. Therefore, Algorithm~\ref{algorithm_bandwidth_allocation_incremental_search} can obtain the optimal solution in the long blocklength regime. In the short-blocklength regime, $r_u^{\Phi, \xi}$ is not concave when $w_k^{\Phi,\xi} \in (0,\infty)$. Nevertheless, based on the results in~\cite{SCJ_drop_packet}, the optimal bandwidth can be obtained in a region $[0,w_{\rm th}] \subset (0,\infty)$, where $r_u^{\Phi, \xi}$ is concave in bandwidth. By searching for the optimal bandwidth in $[0,w_{\rm th}]$, Algorithm~\ref{algorithm_bandwidth_allocation_incremental_search} can obtain the optimal solution in the short blocklength regime\footnote{{The optimality analysis in Section~\ref{section_iterative_algorithm} shows that Algorithm~\ref{algorithm_bandwidth_allocation_incremental_search} can obtain the optimal performance in terms of sum rewards for all three considered QoS requirements in both long and short blocklength regimes. The primary utility of the iterative algorithm lies in serving as a benchmark with optimal reward performance.}}.




\subsection{Feature Engineering and Problem Reformulation}~\label{section_feature_engineering}
To obtain a policy that can generalize well in different scenarios, we propose to use feature engineering technology to represent the channels and QoS requirements with more general features. Specifically, we first normalize the bandwidth allocation policy by the bandwidth reserved for this slice. The normalized bandwidth allocated to the $k$-th user, $k \in \mathcal{K}_{\tau}^{\Phi,\xi}$, is given by ${\tilde{w}_k^{\Phi,\xi}} \triangleq {w}_k^{\Phi,\xi}/W_{\tau}^{\Phi,\xi}$. Then, the normalized minimum bandwidth required by the scheduled users is denoted by $\tilde{\boldsymbol{w}}_{\tau,\min}^{\Phi,\xi} = [\tilde{w}_{1,\min}^{\Phi,\xi}, \tilde{w}_{2,\min}^{\Phi,\xi}, ..., \tilde{w}_{K_{\tau},\min}^{\Phi,\xi}]^{\mathrm{T}}$. We define the surplus bandwidth as
\begin{equation}
    w_{\tau,\mathrm{S}}^{\Phi, \xi} = W_{\tau}^{\Phi,\xi} - \sum_{k \in \mathcal{K}_{\tau}^{\Phi,\xi}} w_k,
    \label{eq_surplus_bandwidth}
\end{equation}
and we further denote the normalized surplus bandwidth by $\tilde{w_{\tau,\mathrm{S}}}^{\Phi,\xi} \triangleq w_{\tau,\mathrm{S}}^{\Phi, \xi}/W_{\tau}^{\Phi,\xi} $.

\begin{figure}[t] 
\centering
\subfigure[{Data flow of our scalable bandwidth allocation algorithm, where $\boldsymbol{H}=\{\boldsymbol{h}_1, \boldsymbol{h}_2, \cdots, \boldsymbol{h}_U\}^{\mathrm{T}}$ is the channel matrix of all the users, and $W_{\tau}^{\Phi,\xi}$ is the reserved bandwidth for the $\tau$-th slice as defined in~eq.~\eqref{eq_reserved_bandwidth}.}]
{\includegraphics[scale=0.8]{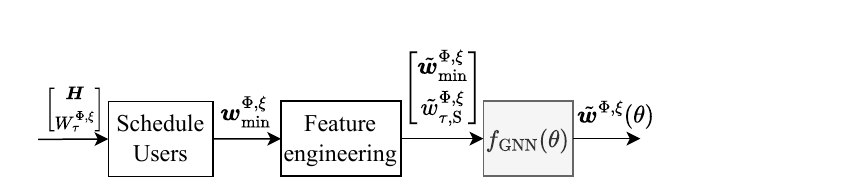}\label{fig_flow_schedule}} 
\subfigure[{Designed GNN for bandwidth allocation.}]
{\includegraphics[scale=0.8]{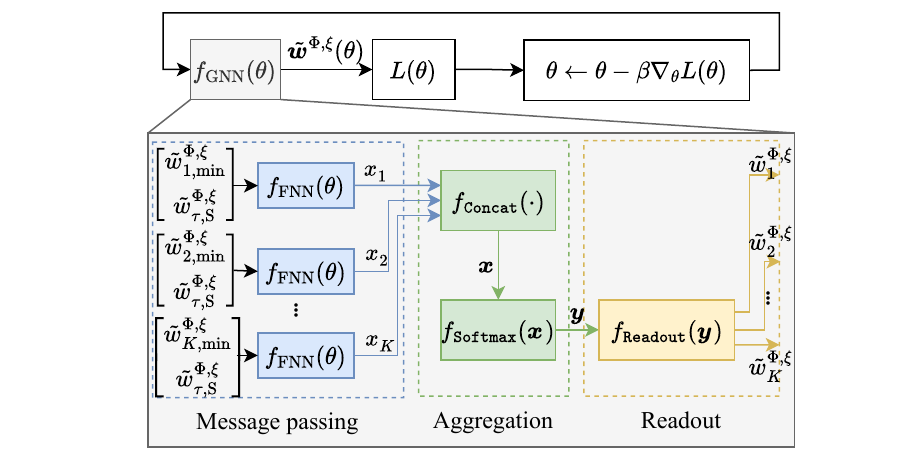}\label{fig_gnn_layer}}
\caption{GNN-based scalable bandwidth allocation.}
\end{figure}

We note that bandwidth allocation policy maps channels and constraints to the bandwidth allocated to each user. After scheduling and normalization, the features of the channel state information and constraints~\eqref{eq_problem_variable_sche_a} and~\eqref{eq_problem_variable_sche_b} can be represented by $\tilde{\boldsymbol{w}}_{\tau,\min}^{\Phi,\xi}$. Therefore, the bandwidth allocation policy can be reformulated as the mapping from $\tilde{\boldsymbol{w}}_{\tau,\min}^{\Phi,\xi}$ and $\tilde{w}_{\tau,\mathrm{S}}^{\Phi,\xi}$ to $\tilde{\boldsymbol{w}}^{\Phi,\xi}$. We denote this function by 
\begin{equation}
    \tilde{\boldsymbol{w}}_{\tau}^{\Phi,\xi} = \boldsymbol{f}^{\rm W}\left(\tilde{\boldsymbol{w}}_{\tau,\min}^{\Phi,\xi},\tilde{w}_{\tau,\mathrm{S}}^{\Phi,\xi}\right)
\end{equation}
where $\boldsymbol{f}^\mathrm{W}(\tilde{\boldsymbol{w}}_{\tau,\min}^{\Phi,\xi},\tilde{w}_{\tau,\mathrm{S}}^{\Phi,\xi}) = \left[f_1^\mathrm{W}(\tilde{\boldsymbol{w}}_{\tau,\min}^{\Phi,\xi},\tilde{w}_{\tau,\mathrm{S}}^{\Phi,\xi}), \cdots, f_K^\mathrm{W}(\tilde{\boldsymbol{w}}_{\tau,\min}^{\Phi,\xi},\tilde{w}_{\tau,\mathrm{S}}^{\Phi,\xi})\right]^\mathrm{T}$ and $\tilde{w}_k^{\Phi,\xi}=f_k^\mathrm{W}(\tilde{\boldsymbol{w}}_{\tau,\min}^{\Phi,\xi}, \tilde{w}_\mathrm{S}^{\Phi,\xi})$. Given the bandwidth reserved for this slice, the achievable rates of the scheduled users can be expressed as
\begin{equation}
\begin{split}
    \boldsymbol{r}_{\tau}^{\Phi,\xi}
    &= 
    \boldsymbol{f}^{\Phi,\xi}\left(\tilde{\boldsymbol{w}}^{\Phi,\xi} \cdot W_{\tau}^{\Phi,\xi}\right)\\
    &= \boldsymbol{f}^{\Phi,\xi}\left(\boldsymbol{f}^\mathrm{W}\left(\tilde{\boldsymbol{w}}_{\tau,\min}^{\Phi,\xi},\tilde{w}_{\tau,\mathrm{S}}^{\Phi,\xi}\right)\cdot W_{\tau}^{\Phi,\xi}\right)
\end{split}
\end{equation}
where $\boldsymbol{r}_{\tau}^{\Phi,\xi}=\left[r_1^{\Phi,\xi}, \cdots, r_{K_{\tau}}^{\Phi,\xi}\right]^\mathrm{T}$, 
$\boldsymbol{f}^{\Phi,\xi}(\tilde{\boldsymbol{w}}_{\tau}^{\Phi,\xi} \cdot W_{\tau}^{\Phi,\xi}) = \left[f_1^{\Phi,\xi}(\tilde{w}_1^{\Phi,\xi} \cdot W_{\tau}^{\Phi,\xi}), \cdots, f_K^{\Phi,\xi}(\tilde{w}_{K_{\tau}}^{\Phi,\xi} \cdot W_{\tau}^{\Phi,\xi})\right]^\mathrm{T}$, 
and $r_k^{\Phi,\xi} = f_k^{\Phi,\xi}\left(f_k^\mathrm{W}(\tilde{\boldsymbol{w}}_{\tau,\min}^{\Phi,\xi}, \tilde{w}_{\tau,\mathrm{S}})\cdot W_{\tau}^{\Phi,\xi}\right)$. 
Then, we can reformulate problem~\eqref{eq_problem_variable_sche} as a functional optimization problem,
\begin{align}
\max_{\boldsymbol{f}^\mathrm{W}(\cdot)}
& \sum\limits_{k \in \mathcal{K}_{\tau}^{\Phi,\xi}} f_k^{\Phi,\xi}\left(f_k^\mathrm{W}\left(\tilde{\boldsymbol{w}}_{\tau,\min}^{\Phi, \xi},\tilde{w}_{\tau,\mathrm{S}}^{\Phi, \xi}\right)\cdot W_{\tau}^{\Phi,\xi}\right), \label{eq_positive_Rs_basic} \\
\mathrm{s.t.}\quad \nonumber
& 1-\sum\limits_{k \in \mathcal{K}_{\tau}^{\Phi,\xi} } f_k^\mathrm{W}\left(\tilde{\boldsymbol{w}}_{\tau,\min}^{\Phi, \xi},\tilde{w}_{\tau,\mathrm{S}}^{\Phi, \xi}\right)\geq 0,         \tag{\ref{eq_positive_Rs_basic}{a}} \label{eq_positive_Rs_basic_a}\\
& f_k^\mathrm{W}\left(\tilde{\boldsymbol{w}}_{\tau,\min}^{\Phi, \xi}, \tilde{w}_{\tau,\mathrm{S}}^{\Phi, \xi}\right) \geq \tilde{w}_{k,\min}^{\Phi, \xi}.  \tag{\ref{eq_positive_Rs_basic}{b}} \label{eq_positive_Rs_basic_b}
\end{align}
In the rest part of this section, we will find the optimal solution to problem~\eqref{eq_positive_Rs_basic}. {The data flow of our scalable bandwidth allocation scheme is shown in Fig.~\ref{fig_flow_schedule}.}

\subsection{Proposed GNN}~\label{section_GNN}
In this subsection, we propose a GNN-based unsupervised learning algorithm to obtain a scalable bandwidth allocation policy.

\begin{algorithm}[t] 
\algsetup{linenosize=\normalsize} \small  
\caption{GNN for Scalable Bandwidth Allocation.} \label{algorithm_bandwidth_allocation_GNN}

Initialize batch size, $J$, number of training epochs, $N$, and learning rate $\beta_{\theta}$.\\
Randomly initialize $\theta^0$.~\label{algorithm_bandwidth_allocation_GNN_line_initial}\\

\For{$n=0, 1, \cdots, N-1$~\label{algorithm_bandwidth_allocation_GNN_line_iter_begin}}
{

Message passing: $x_k^{n} = f_{\mathrm{FNN}}\left(\tilde{w}_{k,\min}^{\Phi,\xi}, \tilde{w}_{\tau,\mathrm{S}}^{\Phi,\xi} \Big| \theta^n\right), \forall k \in \mathcal{K}_{\tau}^{\Phi,\xi}.$~\label{line_message_passing}

Aggregation:
$\boldsymbol{x}^{n}
= f_\texttt{Concat}\left(x_1^{n}, \cdots, x_K^{n}\right)
= \left[x_1^{n}, \cdots, x_K^{n}\right]^\mathrm{T}$ and 
$\boldsymbol{y}^{n} = f_\texttt{Softmax}(\boldsymbol{x}^{n}).$

Readout:      
$\boldsymbol{\tilde{w}}^{n} =f_\texttt{Readout}(\boldsymbol{y}^n) = \boldsymbol{y}^{n} \cdot \tilde{w}_{\tau, \mathrm{S}}^{\Phi,\xi} + \boldsymbol{\tilde{w}}_{\min}^{\Phi,\xi}.$~\label{alg_GNN_line_W}

Update the loss function by eq.~\eqref{eq_lossGNN}, denoted by $f^\mathrm{L}(\theta^{n})$.
~\label{line_loss_GNN}

Update parameters of the GNN by SGA: 
${\theta}^{n+1} ={\theta}^n + \beta_{\theta} {\nabla_{\theta}{f^{\mathrm{L}}}(\theta^{n})}.$~\label{line_SGD_GNN}
}
Return the parameters of the GNN as: $\theta^\mathrm{opt}$.\label{algorithm_bandwidth_allocation_GNN_optimal}
\end{algorithm}

\subsubsection{Structure of GNN}\label{designGNN}
{In our designed GNN, we define the BS and user as two types of vertices in the graph, and each communication link between the BS and a user is represented as an edge in the graph. We note that there are no edges between the users. As shown in Fig.~\ref{fig_gnn_layer}, the proposed GNN structure is based on the message passing neural network (MPNN) structure~\cite{MPNN_chemistry}, and our} GNN-based bandwidth allocation algorithm comprises three key steps {as in an MPNN}: message passing, aggregation, and readout.

\paragraph{Message passing} 
We use a fully connected neural network (FNN) to obtain the embedding of each {user} vertex, denoted by $x_k, \forall k \in \mathcal{K}_{\tau}^{\Phi,\xi}$. The inputs of each FNN include $\tilde{w}_{k,\min}^{\Phi,\xi}$ and $\tilde{w}_{\tau,\mathrm{S}}^{\Phi,\xi} = 1-\sum_{k \in \mathcal{K}_{\tau}^{\Phi,\xi} } \tilde{w}_{k,\min}^{\Phi,\xi}$ {achieved by feature engineering in Section~\ref{section_feature_engineering}}. 
We use $\theta$ to denote the training parameters of the FNN. In the $n$-th epoch, the message passing function, {indicating the message passing from the users to the BS, is given in line~\ref{line_message_passing} of Algorithm~\ref{algorithm_bandwidth_allocation_GNN}}. Since the {user} vertices are homogeneous, the training parameters of the FNN {are reused by the scheduled users. As a result, the number of training parameters of our GNN does not change with the changing numbers of users.}

\paragraph{Aggregation}
In the aggregation step, we first aggregate the embeddings of all the scheduled users by using a concatenation function, $f_\texttt{Concat}(\cdot)$, followed by a $\texttt{Softmax}$ function, $f_{\texttt{Softmax}}(\cdot)$, which serves as the activation function in the aggregation. The output after aggregation is denoted by $\boldsymbol{y}$. {Our GNN can accommodate varying numbers of users, because the output of the number aggregation function matches the count of its inputs as shown in Fig.~1(b), thereby enabling our model's flexibility and adaptability to different user scenarios.}

\paragraph{Readout}
The GNN's output of each vertex is updated by a readout function given by, $f_\texttt{Readout}(\boldsymbol{y}) = \boldsymbol{y} \cdot \tilde{w}_{\tau, \mathrm{S}}^{\Phi,\xi} + \tilde{\boldsymbol{w}}_{\tau,\min}^{\Phi,\xi}$. Since $\boldsymbol{y}$ is obtained from the $\texttt{Softmax}$ function, the summation of its elements is one. From the readout function, all the surplus bandwidth is allocated to the users, and constraints~\eqref{eq_positive_Rs_basic_a} and~\eqref{eq_positive_Rs_basic_b} can be satisfied.



\subsubsection{Unsupervised Learning}
The learning algorithm is detailed in Algorithm~\ref{algorithm_bandwidth_allocation_GNN}. Specifically, in the $n$-th epoch, we use our GNN to obtain the bandwidth allocation and estimate the expectation of the objective function by using the batch samples according to
\begin{align}\label{eq_lossGNN}
f^\mathrm{L}(\theta) = \frac{1}{J}  \sum\limits_{j=1}^{J} \sum\limits_{k \in \mathcal{K}_{\tau}^{\Phi,\xi} } f_{k}^{\Phi, \xi}\left(\tilde{w}_{j,k}^{\Phi,\xi} \cdot W_{j,\tau}^\mathrm{NS}\right),
\end{align}
where $J$ is the batch size. Then, we use stochastic gradient descent (SGA) to maximize the estimated expectation of the objective function in~\eqref{eq_positive_Rs_basic}. As shown in~\cite{SCJ_Book_Chapter}, maximizing the expectation of the objective function, where the expectation is taken over channels, is equivalent to maximizing the objective function with given channels. Thus, from Algorithm~\ref{algorithm_bandwidth_allocation_GNN}, we can find the bandwidth allocation policy that maximizes the objective function in~\eqref{eq_positive_Rs_basic}. 


\subsection{Hybrid-Task Meta-Learning}~\label{section_HML}
To obtain a GNN with strong generalization ability, we propose an HML algorithm that combines multi-task learning and meta-learning.

\subsubsection{Task, Sample, and Taskset}\label{section_definition}
To apply the meta-learning framework, we first define tasks, samples, and tasksets in the context of bandwidth allocation problems. A task is a specific bandwidth allocation problem with a unique combination of system parameters, including the number of users, $U$, the channel model (i.e., path loss model, shadowing, and small-scale channel fading), the QoS requirement, $r_{\tau}^{\Phi,\xi}$, and the reserved bandwidth, $W_{\tau}^{\Phi,\xi}$. If any of the above system parameters change, it would result in a different task. For each task, the samples correspond to the wireless channels that have been transformed into the minimum bandwidth requirement by feature engineering, as specified in constraint~\eqref{eq_positive_Rs_basic_b}. There are four tasksets in meta-learning, and a taskset consists of multiple tasks. We will provide their definitions in the sequel.

\begin{figure}[t]
\centering
%
%
\subfigure[Model-agnostic meta-learning (MAML).]
{\includegraphics[scale=0.8]{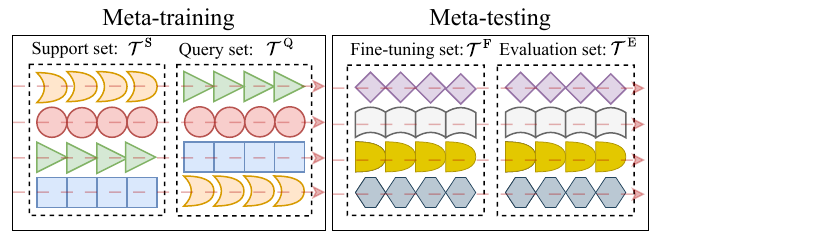}\label{fig_data_distribution_MAML}}
\subfigure[Hybrid-task meta-learning (HML).]
{\includegraphics[scale=0.8]{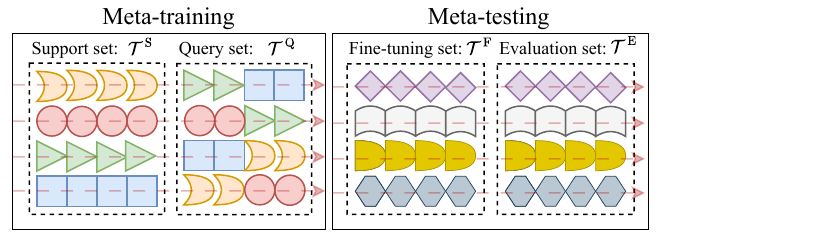}\label{fig_data_distribution_hml}}
\caption{Tasksets of meta-learning algorithms, where different shapes represent different tasks.}
\label{fig_data_distribution}
\end{figure}

\subsubsection{Support Set and Query Set in Meta-Training}
As shown in Fig.~\ref{fig_data_distribution_MAML}, most meta-learning learning algorithms, such as MAML, consist of a meta-training stage and a meta-testing stage. In meta-training, there are two tasksets, support set $\mathcal{T}^\mathrm{S}$ and query set $\mathcal{T}^\mathrm{Q}$. The tasks in the two tasksets are the same, but the samples of each task in the two tasksets are different. Specifically, we first set the initialize parameters of the GNN to $\phi$, which is randomly initialized at the beginning of meta-training, and updated in every iteration of the meta-training. Then, we train the parameters of the GNN by using the tasks and the corresponding samples in the support set, where $\theta$ is initialized with parameters $\phi$. Then, we update the initial parameters $\phi$ by using the tasks and the corresponding samples in the query set. We denote the initial parameters trained in meta-training of MAML by $\phi^*$. The details of the MAML algorithm can be found in~\cite{MAML_Finn}.

\subsubsection{Fine-Tuning Set and Evaluation Set in Meta-Testing} To evaluate the generalization ability of the GNN, a different set of tasks that are unseen in the meta-training stage are used in meta-testing. As shown in Fig.~\ref{fig_data_distribution_MAML}, the tasks in meta-testing are divided into a fine-tuning set and an evaluation set, denoted by $\mathcal{T}^\mathrm{F}$ and $\mathcal{T}^\mathrm{E}$, respectively. The tasks in $\mathcal{T}^\mathrm{F}$ and $\mathcal{T}^\mathrm{E}$ are the same, but the samples of each task in these two tasksets are different. For each new task in meta-testing, the samples from the fine-tuning set are used to fine-tune $\theta$, which is initialized by $\phi^*$ obtained in meta-training. After fine-tuning, the updated GNN is tested with the samples from the evaluation set. If no sample is used to fine-tune the GNN in the meta-testing stage, we refer to this approach as zero-shot meta-learning. Otherwise, it is known as few-shot meta-learning. The meta-testing algorithm is detailed in Algorithm~\ref{algorithm_meta_testing}.

\begin{algorithm}[t] 
\algsetup{linenosize=\normalsize} \small  
\caption{Meta-Testing}
\label{algorithm_meta_testing}
Initialize the number of training epochs, $N$, and the learning rate of the target testing task, $\beta_{\theta}$.\\
{

Select the $i$-th task from the fine-tuning set and the evaluation set: $\mathcal{T}_{i}^\mathrm{F} \in \mathcal{T}^\mathrm{F}$ and $\mathcal{T}_{i}^\mathrm{E} \in \mathcal{T}^\mathrm{E}$.\\
Set the initialization parameters of the GNN as: $\theta^0=\phi^*$.\\
\For{$n=0,1,\cdots,N-1$}
{
Randomly select $J$ samples from task $\mathcal{T}_i^\mathrm{F}$.~\label{line_testing_task_batch_begin}\\
{
\label{line_testing_task_batch_end}}
Calculate loss in the fine-tuning set according to~\eqref{eq_lossGNN}, denoted by $f^\mathrm{L,F}(\theta^{n})$.\\
Update the parameters of GNN by: $\theta^{n+1}=\theta^{n} + {\beta_{\theta}} \nabla_{\theta} {f}^\mathrm{L,F}(\theta^{n})$.\\
}
Randomly select $J'$ samples from task $\mathcal{T}_i^\mathrm{E}$.\\
Evaluate the fine-tuned policies of the $i$-th task using $\theta^{N}$: $\boldsymbol{w}_{i}^{\Phi,\xi}
=f_\mathrm{GNN}\left(\tilde{\boldsymbol{w}}_{\min,i,j'}^{\Phi,\xi}, \tilde{w}_{\mathrm{S},i,j'}^{\Phi,\xi} \big| \theta^{N} \right)$.~\label{alg_meta_testing_query}\\
}
\end{algorithm}

\begin{algorithm}[t] 
\algsetup{linenosize=\normalsize} \small  
\caption{Meta-Training of {Proposed HML}}
\label{algorithm_meta_training}
{
Randomly initialize the training parameters for all the tasks, $\phi$, the number of meta-training epochs, $M$, the learning rate of meta-training, $\beta_{\phi}$, and the learning rate of each task, $\beta_{\theta}$.\\
\For {$m=0,1,\cdots,M-1$}
{
Select a batch of $I$ tasks from the support set: $\mathcal{T}_i^\mathrm{S} \in \mathcal{T}^\mathrm{S}, \quad i \in \{1, 2, \cdots, I\}$.\\
\For {$i = 1,2, \cdots, I$}
{
Set the initial parameters of the GNN to
$\theta_{i}^m=\phi^m.$\\
\For{$n=0,1,\cdots,N-1$}
{
Randomly select $J$ samples from task $\mathcal{T}_i^\mathrm{S}$.\\
{

Calculate the loss function in the support set according to~\eqref{eq_lossGNN}, denoted by ${f}^\mathrm{L,S}(\theta_{i}^{m,n})$.~\label{alg_meta_training_support_loss}

Update the parameters by:
$\theta_{i}^{m,n+1}=\theta_{i}^{m,n} + {\beta_{\theta}} \nabla_{\theta} {f}^\mathrm{L,S}(\theta_i^{m,n}).$~\label{alg_meta_training_support_SGD}
}
}

Select a batch of $I'$ tasks from the query set: $\mathcal{T}_{i'}^\mathrm{Q} \in \mathcal{T}^\mathrm{Q}, \quad i' \in \{1, 2, \cdots, I'\}$.


\For{$i'=1,\cdots,I'$}
{
Randomly select $J'$ samples from task $\mathcal{T}_{i'}^\mathrm{Q}, \quad j' \in \{1,2,...,J'\}$.\\
Calculate the loss function in the query task:
$f_{i}^\mathrm{L,Q}\left(\theta_{i}^{m,N}\right)
= \frac{1}{I'} \frac{1}{J'} \sum\limits_{i'=1}^{I'} \sum\limits_{j'=1}^{J'} \sum\limits_{k \in \mathcal{K}_{\tau}^{\Phi,\xi} } f_k^{\Phi, \xi}\left(\tilde{w}_{i',j',k}^{\Phi,\xi,m} \cdot W_{\tau,j}^\mathrm{NS}\right)$,
~\label{alg_loss_individual_testing_task}
}
}
Calculate the loss function in meta-training: ${f}^{\mathrm{L,Meta},m}\left(\phi^{m}\right) = \frac{1}{I}\sum\limits_{i=1}^{I} {f}_i^\mathrm{L,Q}\left(\theta_i^{m,N}\right)$.~\label{alg_loss_meta}

Update the initial parameters:
$\phi^{m+1} = \phi^{m} + \beta_{\phi} \nabla_{\phi} {f}^{\mathrm{L,Meta},m}\left(\phi^{m}\right)$.\\
}
Return the optimal initial parameters of the GNN: $\phi^\mathrm{opt} = \phi^{M}$.\label{algorithm_bandwidth_allocation_HML_optimal}
}
\end{algorithm}

\subsubsection{Meta-Training of Proposed HML Algorithm}
Fig.~\ref{fig_data_distribution_hml} illustrates the tasks and tasksets used in the meta-training and meta-testing of the proposed HML algorithm. The difference between MAML and HML lies in the selection of tasks from the query set. In MAML, the tasks selected from the query set are identical to those selected from the support set in each meta-training epoch. To improve the generalization ability, in HML, we select different tasks from the query set to train the initial parameters of the GNN. Specifically, $I'$ tasks are randomly selected from the query set to estimate the average loss of the GNN parameterized by $\phi^m$ in the $m$-th epoch of meta-training. 
{In the meta-training stage, our GNN learns the hidden features across the diverse tasks, thus facilitating the adaptation ability in novel tasks in the meta-testing stage.}
The step-by-step algorithm for meta-training of the proposed HML algorithm is described in Algorithm~\ref{algorithm_meta_training}, and the meta-testing algorithm of HML is the same as that of MAML in Algorithm~\ref{algorithm_meta_testing}.

\subsection{{Complexity Analysis}}
\subsubsection{{Computational Complexity Derivations}}
{We compare the computational complexity of our GNN and HML algorithms with the iterative algorithm given in Algorithm~\ref{algorithm_bandwidth_allocation_incremental_search}. In our considered cellular systems, all the algorithms will be implemented in each transmission time interval with a duration of less than $1$ ms. Thus, we are interested in the inference complexity of the proposed GNN and HML algorithms, i.e., the number of operations to be executed to obtain the bandwidth allocation in each transmission time interval.
}
%
%

To compute the embedding of each vertex in the GNN, we first calculate the output of the FNN depicted in Fig.~\ref{fig_gnn_layer}. We denote the number of layers of the FNN by $L_\mathrm{FNN}$ and the number of neurons in the $\ell$-th layer by $m_\mathrm{FNN}^{\ell}$. Then, the number of multiplications required to compute the output of the $\ell$-th layer is $m_\mathrm{FNN}^{\ell} \cdot m_\mathrm{FNN}^{\ell+1}$, and the total number of multiplications for computing the embedding is $M_\mathrm{FNN} = \sum\nolimits_{\ell=1}^{L_\mathrm{FNN}} m_\mathrm{FNN}^{\ell} \cdot m_\mathrm{FNN}^{\ell+1}$~\cite{SCY_transfer}. After obtaining the embeddings of the $K$ {scheduled} users {with sufficiently strong channels}, the number of multiplications required by $f_\texttt{Softmax}(\boldsymbol{x})$ and $f_\texttt{Readout}(\boldsymbol{y})$ is $2K$. Therefore, the inference complexity of the GNN-based bandwidth allocation policy is
{
\begin{equation}
O_\mathrm{GNN}=O( \Omega_{\mathrm{US}} + K \cdot (M_\mathrm{FNN}+2)),
\label{eq_complexity_GNN}
\end{equation}
where $\Omega_{\mathrm{US}}$ represents the computational complexity for the user scheduling process. Since HML employs the same neural network architecture as our GNN, the inference latency of HML is equivalent to that of the GNN and is given by
\begin{equation}
O_\mathrm{HML} = O_\mathrm{GNN}.
\label{eq_complexity_HML}
\end{equation}
For the iterative algorithm, as shown in line~\ref{iterative_DeltaW} of Algorithm~\ref{algorithm_bandwidth_allocation_incremental_search}, in} each iteration, we assign a small portion of the normalized surplus bandwidth, denoted by $\Delta \tilde{w}$, to a user that can maximize the objective function. The algorithm needs to compute the objective function $K$ times {to} find the best user {in each iteration}. We denote the complexity for computing the objective function by $\Omega^{\Phi,\xi}$
, then the complexity of the iterative algorithm is given by
\begin{equation}
{
O_\mathrm{iter}=O\left(\Omega_{\mathrm{US}} + K \cdot\frac{ w_{\tau, \mathrm{S}}^{\Phi,\xi} }{\Delta w} \cdot  \Omega^{\Phi,\xi}\right),
}
\label{eq_complexity_iterative}
\end{equation}
{where $w_{\tau, \mathrm{S}}^{\Phi,\xi}$ is the surplus bandwidth defined in eq.~\eqref{eq_surplus_bandwidth}, and $w_{\tau, \mathrm{S}}^{\Phi,\xi}/{\Delta {w}}$} represents the number of iterations used in the iterative algorithm.
%
%
\subsubsection{{Complexity Comparative Analysis}\label{section_complexity_compare_analysis}}
To obtain bandwidth allocation in each transmission time interval, the transmitter either uses the forward propagation algorithm to compute the outcome of the {GNN/HML} or executes the iterative algorithm. From eqs.~\eqref{eq_complexity_GNN} and~\eqref{eq_complexity_iterative}, we can see that the computational complexity of our {GNN/HML} and the iterative algorithm increase linearly with the number of {scheduled users, $K$. Recall that $M_\mathrm{FNN}$ in eq.~\eqref{eq_complexity_GNN} is a small number~{\cite{SCY_transfer}}, in contrast, although the complexity of the iterative algorithm also depends on the channels of the scheduled users, it increases with the amount of surplus bandwidth, $w_\mathrm{S}$, and decreases with the resource block, $\Delta w$. More importantly, the computing complexity for evaluating the objective function, $\Omega^{\Phi,\xi}$, in each iteration of the optimization algorithm could also be extremely high. Therefore}, the inference complexity of the GNN is much lower than the complexity of the iterative optimization algorithm.
%
%
%

\begin{table}[t] 
\renewcommand\arraystretch{1} 
\caption{Key Simulation Parameters} 
\centering 
\begin{tabular}{l l}
\toprule 
\toprule 
\textbf{Simulation parameters}                                          &\textbf{Values}\\
\midrule
Transmit power of each user, $P_u$                                      &23 dBm\\
\hline
Single-sided noise spectral density, $N_0$                              &-174 dBm/Hz\\
\hline
Channel coherence time, $T_\mathrm{c}$                                  &$1$ms~\cite{SCY_Rs_short}\\
\hline
Duration of one time slot, $T_\mathrm{s}$                               &$0.125$ms\\
\hline
Decoding error probability, $\epsilon_{u}$                              &$10^{-5}$~\cite{SCY_Rs_short}\\
\hline
Information leakage, $\delta_{u}$                                       &$10^{-2}$~\cite{SCY_Rs_short}\\
\hline
QoS exponent of queuing delay, $\vartheta_{u}$                          &$10^{-3}$~\cite{SCY_Rs_short}\\
\hline
Size of bandwidth resource block, $\Delta {w}$                          &$10$~kHz\\
\hline
Learning rates, $\beta_{\theta}/\beta_{\phi}$                           &$10^{-4}$\\ 
\hline
Batch sizes of GNN, $J/J'$                                              &32\\ 
\hline
Batch sizes of meta optimizer, $I,I'$                                   &4, 2\\ 
\bottomrule 
\bottomrule 
\end{tabular} 
\label{table_Simulation_Parameters} 
\end{table}

\begin{table*}[t] 
\renewcommand\arraystretch{1} 
\caption{System Parameters of Different Tasks} 
\label{table_Simulation_task_parameters}
\centering 
\begin{tabular}{l | l | l | l}
\toprule 
\toprule 
     &{Parameters} &$\mathcal{T}^{\mathrm{S}}$ \& $\mathcal{T}^{\mathrm{Q}}$                &$\mathcal{T}^{\mathrm{F}}$ \& $\mathcal{T}^\mathrm{E}$\\
\midrule
Network scale                                             
&Number of users                                          &$U_{\tau}^{\Phi,\xi}\in \{10, 11, \cdots, 30\}$                   &$U_{\tau}^{\Phi,\xi}=50$\\
\hline
\multirow{3}{*}{Channel models} &Path loss: $\alpha_u = (d_u)^{-\gamma_u}$       &$\gamma_u \in \{2,3\} $                                         &$\gamma_u=4$\\
\cline{2-4}
&\makecell[l]{
Shadowing:\\ 
$p_u^\mathrm{S}(\psi) = \frac{10/\ln10}{\sqrt{2 \pi} \sigma_{\psi_\mathrm{dB}} \psi} 
\exp\left(-\frac{(10\log_{10}\psi - \mu_{\psi_\mathrm{dB}})^2}{2\sigma_{\psi_\mathrm{dB}}^2}\right)$  }                               &\makecell[l]
{$\psi_\mathrm{dB}\in \{3,4,5\}$}                               &$\psi_\mathrm{dB}=8$\\
\cline{2-4}
\rule{0pt}{15pt}
&\makecell[l]{
Small-scale channels:   
\\
$p_u^\mathrm{I}(z| s,\sigma) = \frac{z}{\sigma^2}\exp\left(-\frac{z^2+s^2}{2 \sigma^2}\right) \cdot I_0\left(\frac{zs}{\sigma^2}\right)$,
\\
$p_u^\mathrm{N}(z|m, \sigma) =\frac{2m^m z^{2m-1}}{\Gamma(m) {(2\sigma^2)}^m} \exp\left(-\frac{m z^2}{2\sigma^2} \right)$, 
\\
$p_u^\mathrm{R}(z| \sigma)=\frac{z}{\sigma^2}\exp\left(-\frac{z^2}{2 \sigma^2}\right)$
}                                  
&\makecell[l]
{$p_u^\mathrm{I}(z| s,\sigma), s\in \{1\cdots 5\}$, \vspace{1mm} 
\\  
$p_u^\mathrm{N}(z|m, \sigma), m\in \{2, \cdots, 6\}$
}    &$p_u^\mathrm{R}(z| \sigma)$
\\
\hline
\multirow{2}{*}{QoS}                
&Rewards with different QoS requirements                                 &$\max\limits_{\boldsymbol{w}}   \sum\limits_{u \in \mathcal{U}_{\tau}^{S,\mathcal{I}}} r_u^{S,\mathcal{I}}$  
&\thead[l]{
$\max\limits_{\boldsymbol{w}} \sum\limits_{u \in \mathcal{U}_{\tau}^{\Phi,\xi}} r_u^{\Phi,\xi}$, 
\vspace{1mm} 
\\
$\Phi \in \{D, E, S\}$,
$\xi \in \{\mathcal{I,F}\}$
}
\\
\cline{2-4}
\rule{0pt}{10pt} 
&Values of QoS constraints (Mbps)                                  &$r_{\tau}^{S,\mathcal{I}} \in \{1, \cdots, 10\}$                  &\makecell[l]{
\vspace{1mm} 
$r_{\tau}^{\Phi,\xi} = 10$
}
\\
\hline
\rule{0pt}{9pt} 
Reserved bandwidth &Constraints on reserved bandwidth (MHz)                       &$W_{\tau}^{S,\mathcal{I}} \in \{10, \cdots, 100\}$                &$W_{\tau}^{\Phi,\xi}=100$
\\
\bottomrule
\bottomrule
\end{tabular} 
\end{table*}

\section{Performance Evaluation}
In this section, we evaluate the performance of our GNN-based HML algorithm. The GNN is first initialized by the parameters obtained from meta-training, where all the tasks aim to maximize the sum of the secrecy rate with different numbers of users and channel models. Then, we evaluate the performance of our GNN in unseen tasks with different numbers of users, channel models, objective functions, QoS constraints, and reserved bandwidth.



\subsection{System Setup}~\label{SimSetup}
We consider a BS, located at $(0,0)$~m, serving multiple users randomly distributed in a rectangular area, where the coordinates of the users are denoted by $(c_{x,u},c_{y,u})$, where $c_{x,u}$ and $c_{y,u}$ $\in [-100,100]$. When the QoS requirement is secrecy rate, an eavesdropper is randomly located in the above rectangular area. The transmitted signal of each user is a complex Gaussian process with zero-mean and equal variance, $\sigma^2=1$. Channel models include large-scale channels and small-scale channels. Specifically, the large-scale channels depend on path loss and shadowing fading, whilst small-scale channels follow Rice, Nakagami, and Rayleigh distributions with various parameters in Table \ref{table_Simulation_task_parameters}. The number of neurons in each layer of the GNN is $2/32/64/32/1$. Unless otherwise mentioned, the simulation parameters are summarized in Table~\ref{table_Simulation_Parameters}, and the parameters of tasksets are defined in Table~\ref{table_Simulation_task_parameters}. 
{In real-world communication systems, the BSs capture the information changes in the environment according to the 5G New Radio standard~\cite{5G_NR}.}

\subsection{Performance of GNN}
\begin{figure}[t]
\centering
{\includegraphics[width=9cm, height=6.3cm]{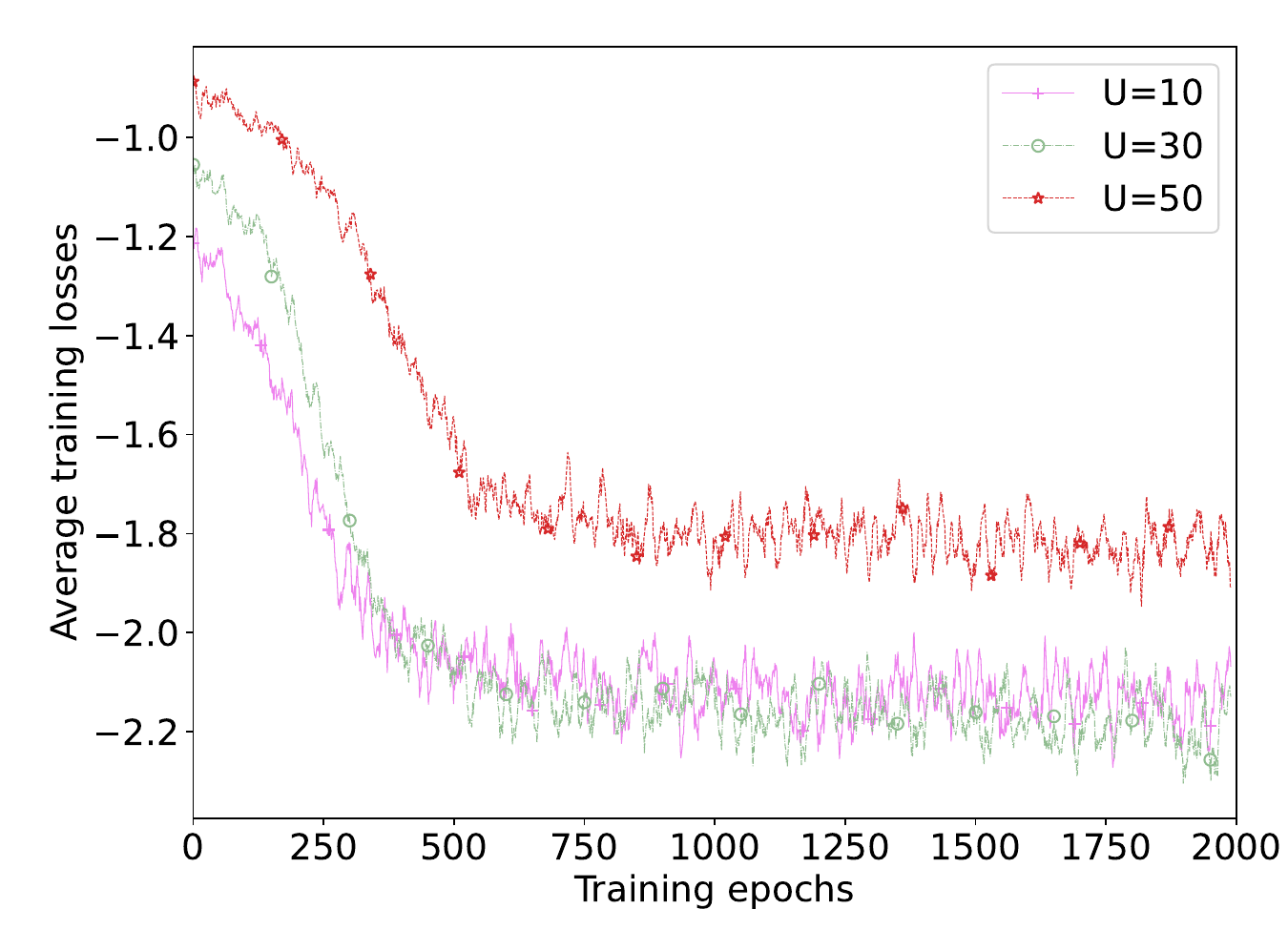}}
\caption{{Average t}raining losses with different numbers of users, where the secrecy rate in the long blocklength regime is considered, $r_{\tau}^{S,\mathcal{I}}=10$~Mbps, and $W_{\tau}^{S,\mathcal{I}} = 100$~MHz.}
\label{fig_training_scalable_GNN}
\end{figure}
%
Fig.~\ref{fig_training_scalable_GNN} shows the training losses when the number of users increases from $10$ to $50$. The results show that the unsupervised learning algorithm can converge after a few hundred training epochs for different numbers of users, and the convergence time increases slightly with the number of users.

\begin{figure}[!t]
    \centering
    \subfigure[Secrecy rates of scheduled users.]
    {\includegraphics[width=9cm, height=6.3cm]{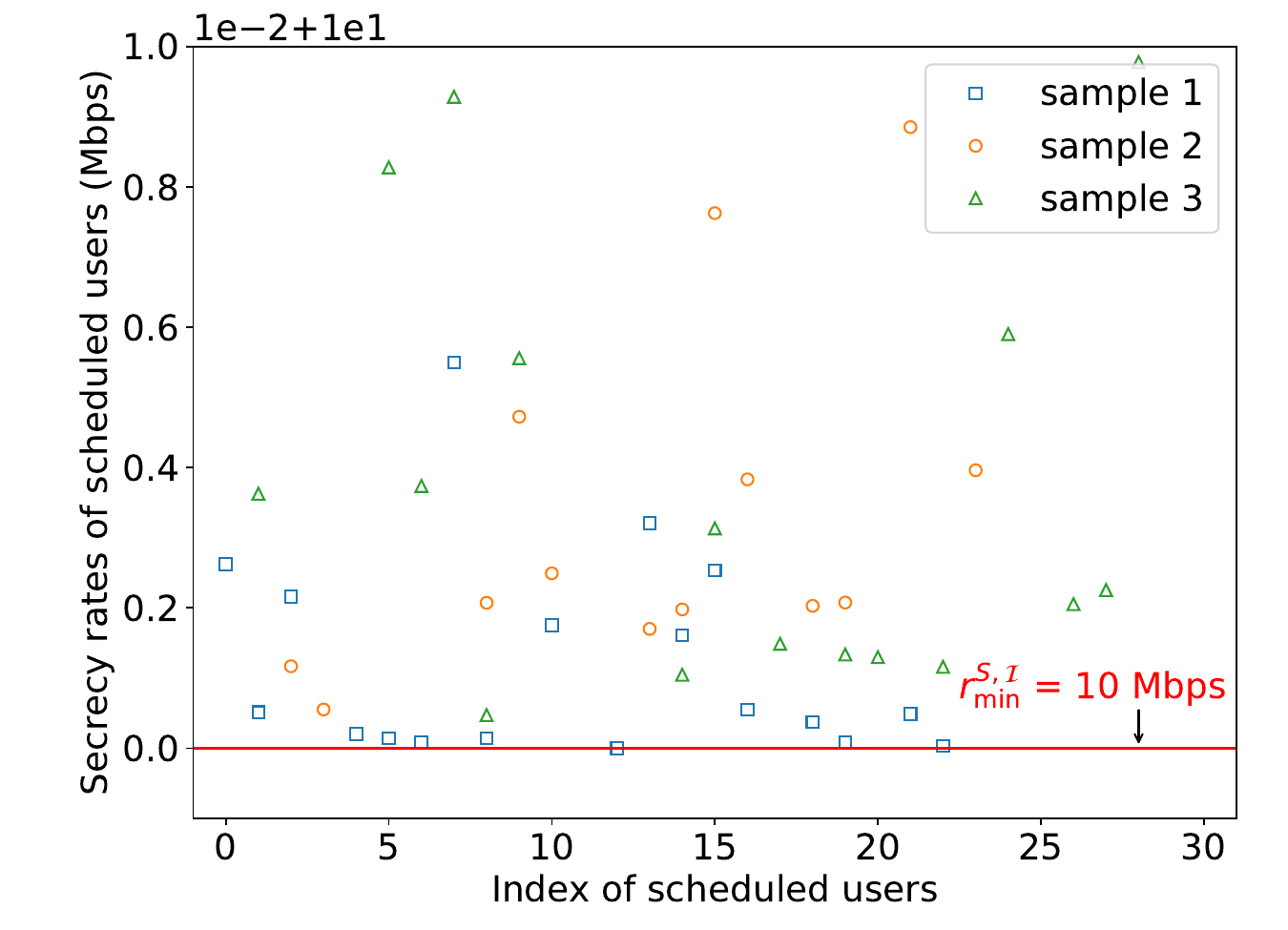}\label{fig_satisfy_Rs_min}}
    \subfigure[Sum secrecy rate.]
    {\includegraphics[width=9cm, height=6.3cm]{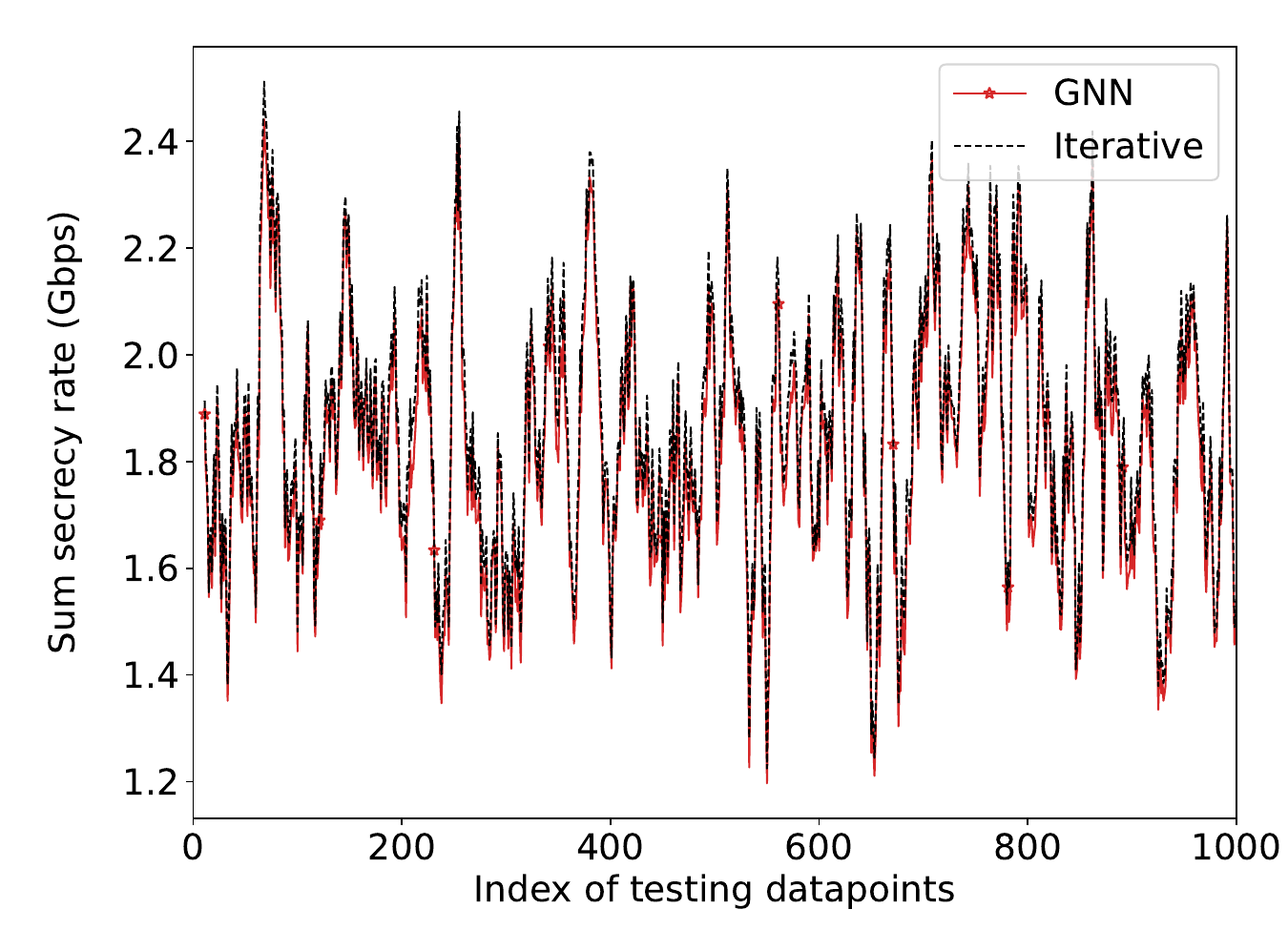}\label{fig_RsSum_test_GNN}}
    \caption{Testing samples are selected from taskset $\mathcal{T}^\mathrm{F}$ and $\mathcal{T}^\mathrm{E}$ in Table.~\ref{table_Simulation_task_parameters}.}
    \label{fig_sim_test_GNN}
\end{figure}
After the training stage of the unsupervised learning algorithm, we select $1000$ samples from the evaluation set of the same task to evaluate the constraint and reward achieved by the GNN in Fig.~\ref{fig_sim_test_GNN}. The results in Fig.~\ref{fig_satisfy_Rs_min} show that the secrecy rates of all the scheduled users are equal to or higher than the requirement, $r_{\tau}^{S,\mathcal{I}}=10$~Mbps. The results in Fig.~\ref{fig_RsSum_test_GNN} show that the sum secrecy rate achieved by the GNN is close to that achieved by the iterative optimization algorithm in Section~\ref{section_iterative_algorithm} (with legend ``{Iterative}''). In other words, the unsupervised learning algorithm can obtain a near-optimal solution.

\subsection{Meta-Testing Performance of HML}
In this subsection, we evaluate the generalization ability of the proposed HML algorithm. The differences between tasks in meta-training and meta-testing are shown in  Table.~\ref{table_Simulation_task_parameters}. {In meta-training, we randomly select the parameters included in $\mathcal{T}^\mathrm{S}$ and $\mathcal{T}^\mathrm{S}$ in each training epoch.} In meta-testing, we first select an unseen task that is not included in meta-training. In each training epoch of the meta-testing, $32$ samples are randomly selected from $\mathcal{T}^\mathrm{F}$ to fine-tune the GNN, whilst all the $1000$ testing samples from the same task in $\mathcal{T}^\mathrm{E}$ are used to evaluate the performance. 

\subsubsection{Different Wireless Channels and QoS Requirements}
\begin{figure}[!t]
\centering 
\subfigure[Secrecy rate in long blocklength regime.]
{\includegraphics[width=9cm, height=6.3cm]{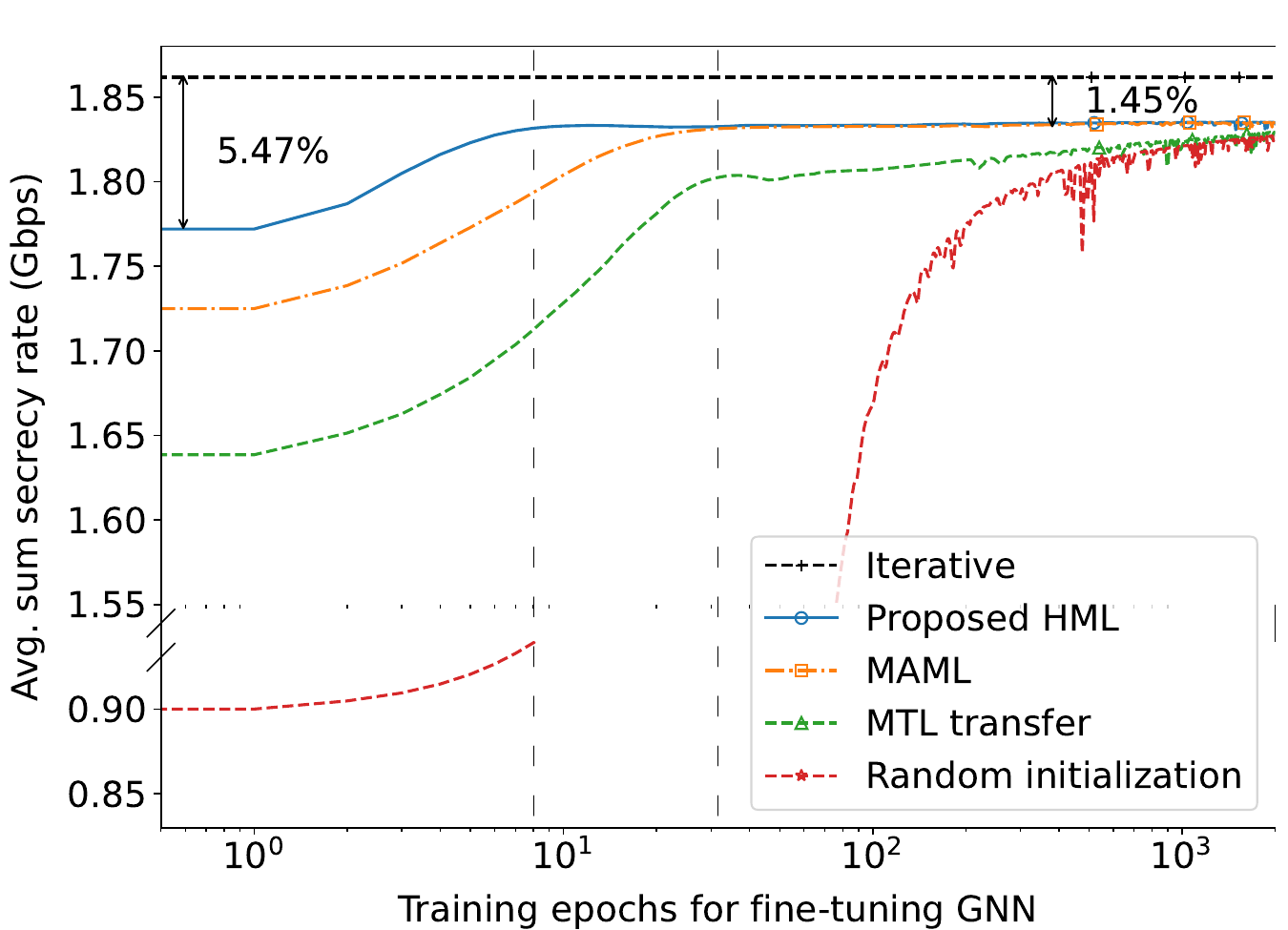}\label{fig_maximize_L_sum_secrecy_rate}} 
\subfigure[Secrecy rate in short blocklength regime.]
{\includegraphics[width=9cm, height=6.3cm]{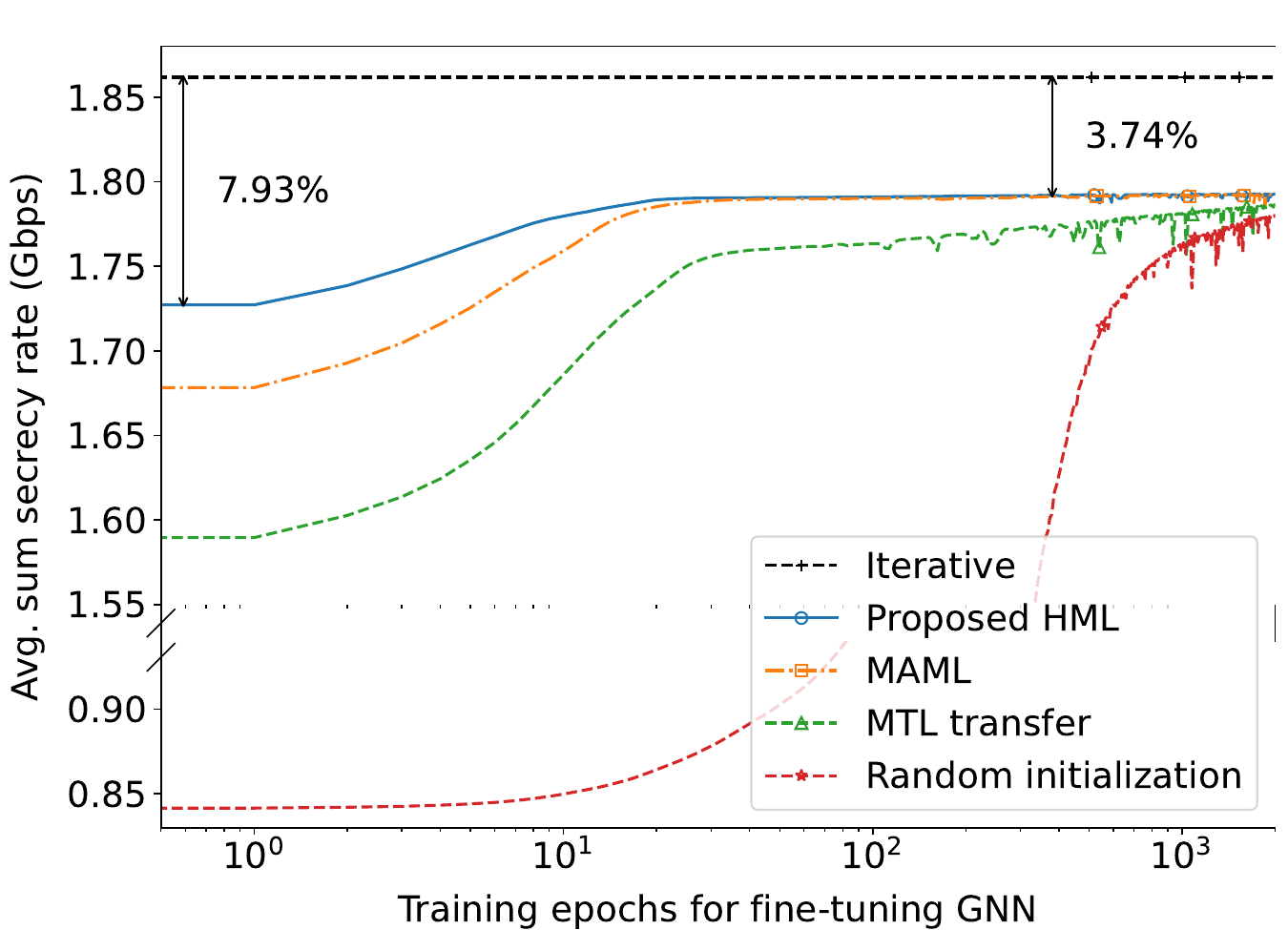}\label{fig_maximize_S_sum_secrecy_rate}} 
\caption{Meta-testing with unseen channel models.}
\label{fig_sim_objective_secrecy_rate}
\end{figure}

\begin{figure}[!t]
\centering 
\subfigure[Shannon capacity in long blocklength regime.]
{\includegraphics[width=9cm, height=6.3cm]{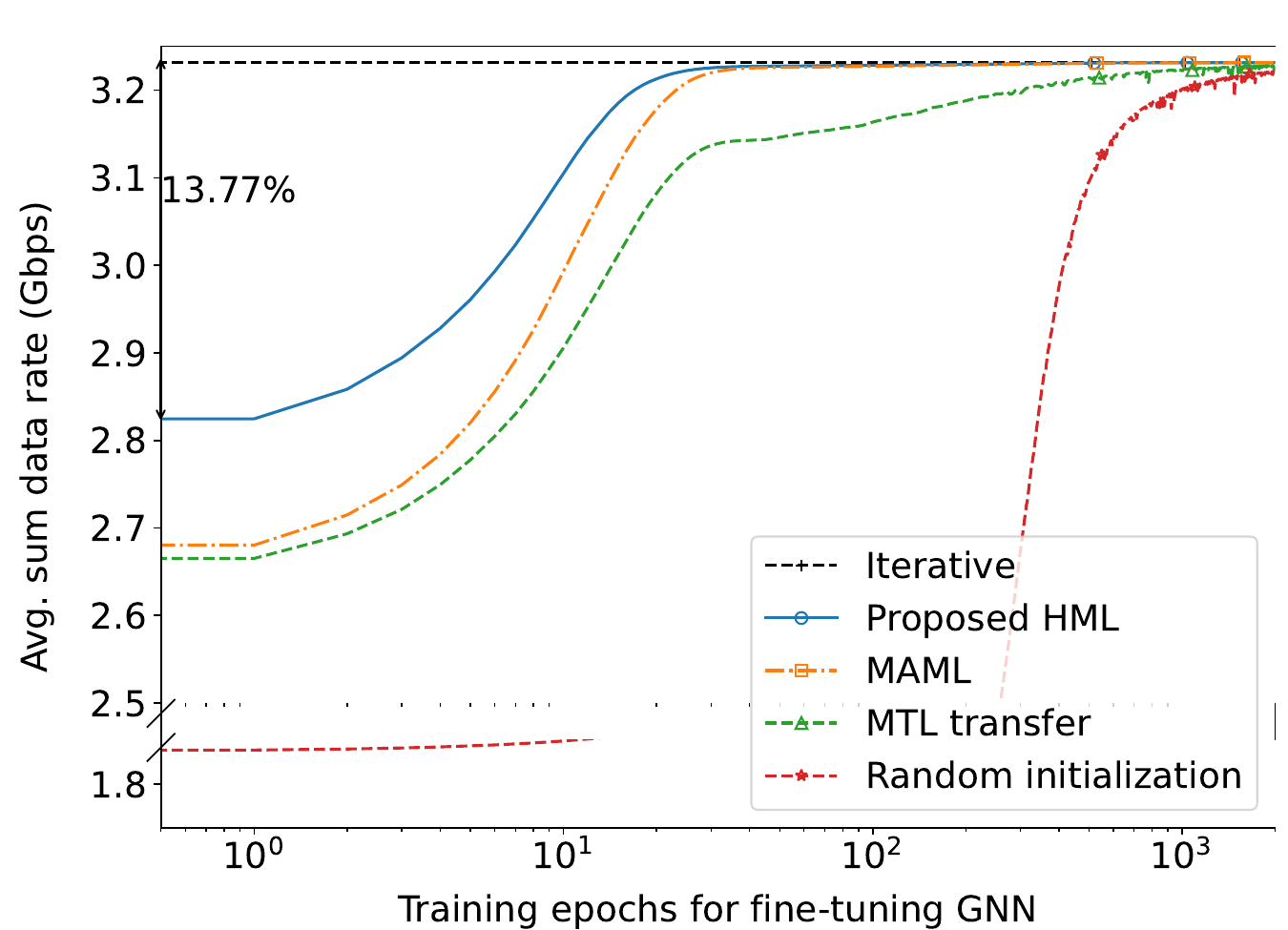}\label{fig_maximize_L_sum_data_rate}} 
\subfigure[Achievable rate in short blocklength regime.]
{\includegraphics[width=9cm, height=6.3cm]{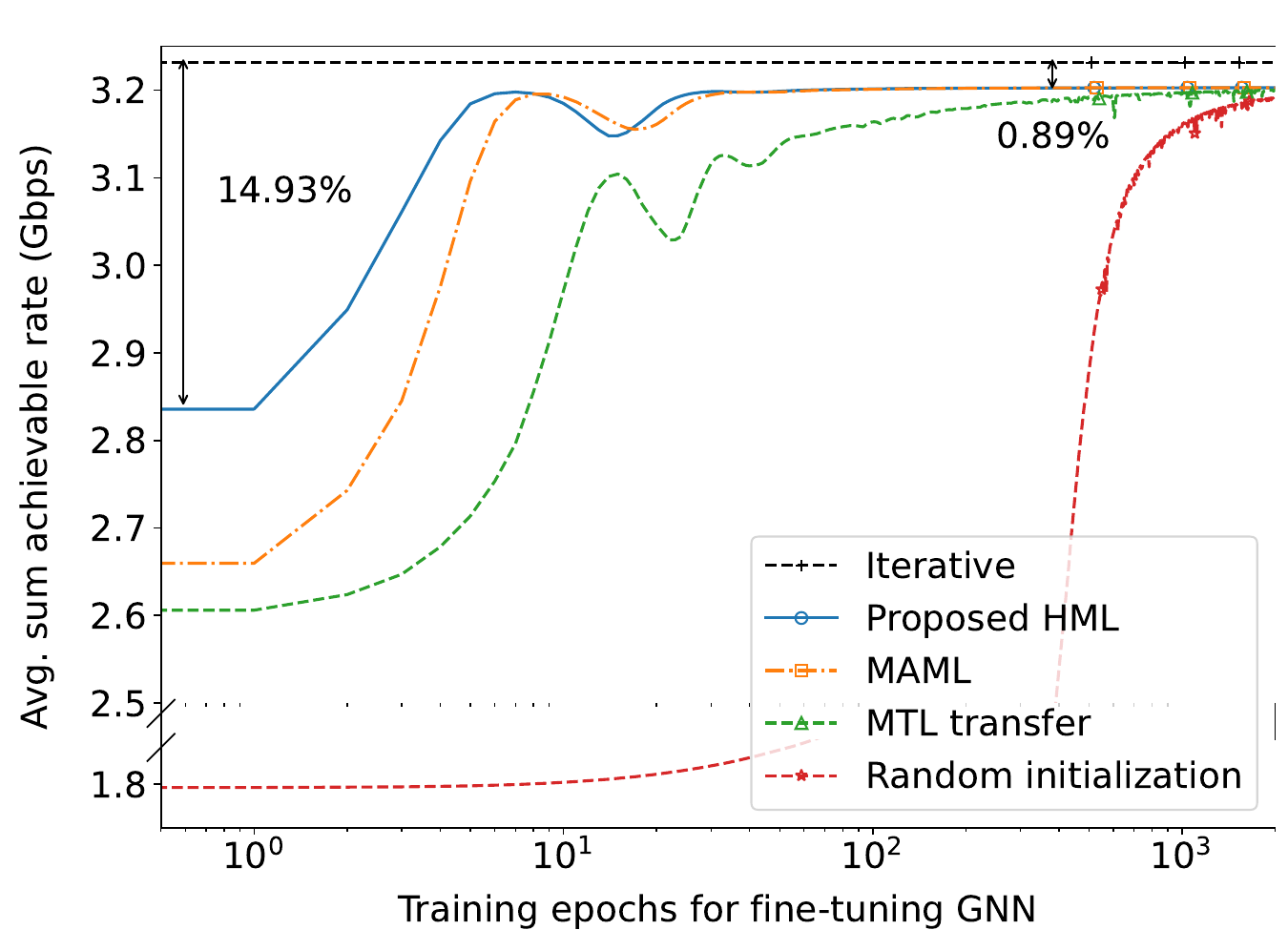}\label{fig_maximize_S_sum_data_rate}} 
\caption{Meta-testing with unseen QoS requirements of rate rates and unseen channels.}
\label{fig_sim_objective_data_rate}
\end{figure}

\begin{figure}[!t]
\centering 
\subfigure[Effective capacity in long blocklength regime.]
{\includegraphics[width=9cm, height=6.3cm]{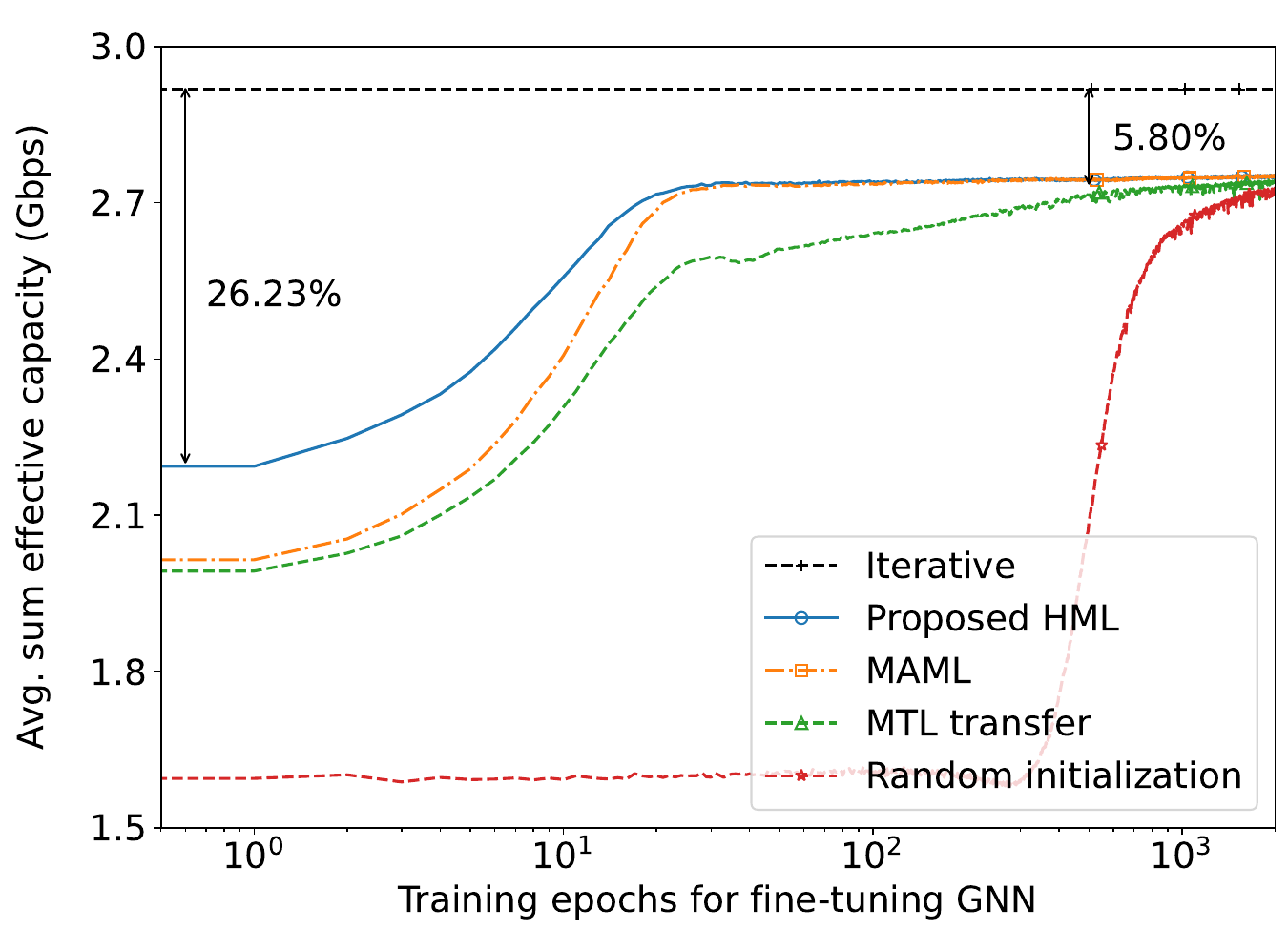}\label{fig_maximize_L_sum_effective_rate}} 
\subfigure[Effective capacity in short blocklength regime.]
{\includegraphics[width=9cm, height=6.3cm]{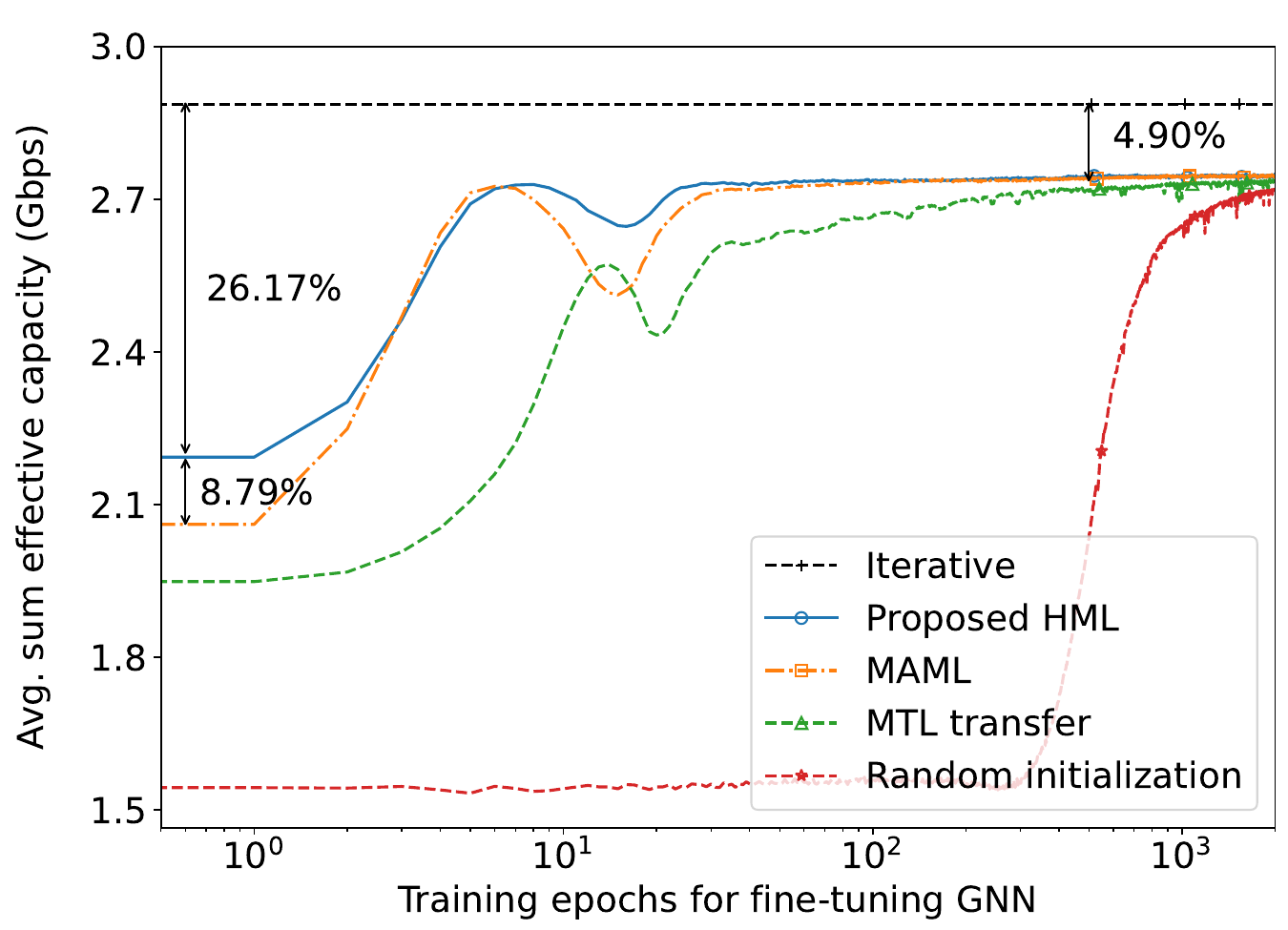}\label{fig_maximize_S_sum_effective_rate}} 
\caption{Meta-testing with unseen QoS requirements and unseen channels.}
\label{fig_sim_objective_effective_capacity}
\end{figure}

In this part, we set $W_{\tau}^{S,\mathcal{I}}=100$~MHz and $r_{\tau}^{S,\mathcal{I}}=10$~Mbps for all types of services. The other parameters follow the rules in $\mathcal{T}^\mathrm{S}$ and $\mathcal{T}^\mathrm{Q}$ as shown in Table.~\ref{table_Simulation_task_parameters}. We compare the initial performance and sample efficiency of HML with four benchmarks: 1) {Iterative}, 2) Model-agnostic meta-learning (MAML), 3) Multi-task learning-based transfer learning (MTL Transfer), and 4) Random initialization.
\begin{itemize}
    \item \textit{{Iterative}}: The optimal solution is obtained by the iterative algorithm detailed in Section~\ref{section_iterative_algorithm}. 
    %
    \item \textit{MAML}: MAML is one of the most widely used meta-learning algorithms, and its key ideas have been discussed in Section~\ref{section_HML}.
    \item \textit{MTL Transfer}: Transfer learning improves the sample efficiency by fine-tuning the parameters of the pre-trained GNN in a task with fewer training samples. With multi-task learning (MTL), the initial performance is much better than random initialization as the GNN is pre-trained in multiple tasks~\cite{MAML_Finn, TWC_MTL_NOMA}. To execute MTL transfer learning, we only need to replace the initialization in line ~\ref{algorithm_bandwidth_allocation_GNN_line_initial} of Algorithm~\ref{algorithm_bandwidth_allocation_GNN} by the pre-trained parameters. 
    \item \textit{Random Initialization}: Random initialization is the conventional method that trains the GNN from scratch with a new task.
\end{itemize}
In figures~\ref{fig_sim_objective_secrecy_rate}-\ref{fig_sim_objective_effective_capacity}, the horizontal axis represents the training epochs used to fine-tune the GNN, and $32$ samples from $\mathcal{T}^{\rm F}$ are used to train the GNN. The vertical axis represents the sum of the rewards of all the users, and the average is taken over samples, i.e., $1000$ testing samples from $\mathcal{T}^{\rm E}$ are used. We refer to it as the average sum reward.

In Fig.~\ref{fig_sim_objective_secrecy_rate}, we consider the average sum of secrecy rates and illustrate the impacts of the number of users, channel models, and coding blocklength on the initial performance and sample efficiency of different methods. 
The results in Fig.~\ref{fig_sim_objective_secrecy_rate} show that HML achieves the best initial average sum secrecy rate and the highest sample efficiency compared with all the benchmarks. {In Fig.~\ref{fig_maximize_L_sum_secrecy_rate}, HML can converge in $8$ training epochs. Both MAML and MTL transfer learning take more than $30$ epochs to converge. Thus, HML can reduce the convergence time by up to $73\%$.} After the fine-tuning, the gap between learning methods and the optimal solution is around $1.45$\%. In Fig.~\ref{fig_maximize_S_sum_secrecy_rate}, the coding blocklength in meta-testing is also different from that in meta-training. As a result, the gap between the initial performance of HML and the optimal solution is $7.93\%$. After fine-tuning, the gap reduced to $3.74\%$, which is larger than the gap in Fig.~\ref{fig_maximize_L_sum_secrecy_rate}, where the blocklength is the same in meta-training and meta-testing.


Fig.~\ref{fig_sim_objective_data_rate} shows the average sum of data rates achieved by different methods. The results indicate that when the reward function and the QoS constraint in meta-testing are different from that in meta-training, the gaps between the initial performance of HML and the optimal solution increase to $13.77\%$ and $14.93\%$ in long and short blocklength regimes, respectively. After fine-tuning, the gaps between the learning methods and the optimal solution are smaller than that in Fig.~\ref{fig_sim_objective_secrecy_rate}. This is because Shannon's capacity/achievable rate are two special cases of the secrecy rate in the long/short blocklength regimes when the {eavesdrop}ped channels are in deep fading. It is easier to learn a good policy when the problem becomes less complicated.


Fig.~\ref{fig_sim_objective_effective_capacity} shows the average sum of effective capacities achieved in the meta-testing stage, where the initial parameters of the GNN are obtained from meta-training, and the GNN is trained with tasks maximizing the sum secrecy rate in the long blocklength regime. In other words, the QoS requirement in meta-testing is queuing delay requirement, which is quite different from the security requirement in meta-training. By comparing the results in Figs.~\ref{fig_sim_objective_effective_capacity} and~\ref{fig_sim_objective_secrecy_rate}, we can observe that the gaps between the HML and the optimal solution in Fig.~\ref{fig_sim_objective_effective_capacity} are larger than the gaps in Fig.~\ref{fig_sim_objective_secrecy_rate}. Nevertheless, HML can still converge in around $10$ to $30$ epochs and outperforms the other benchmarks in Fig.~\ref{fig_sim_objective_effective_capacity}. 

\begin{figure}[!t]
\centering 
\subfigure[$r_{\tau}^{S,\mathcal{I}}=10$~Mbps in meta-training.] 
{\includegraphics[width=9cm, height=6.3cm]{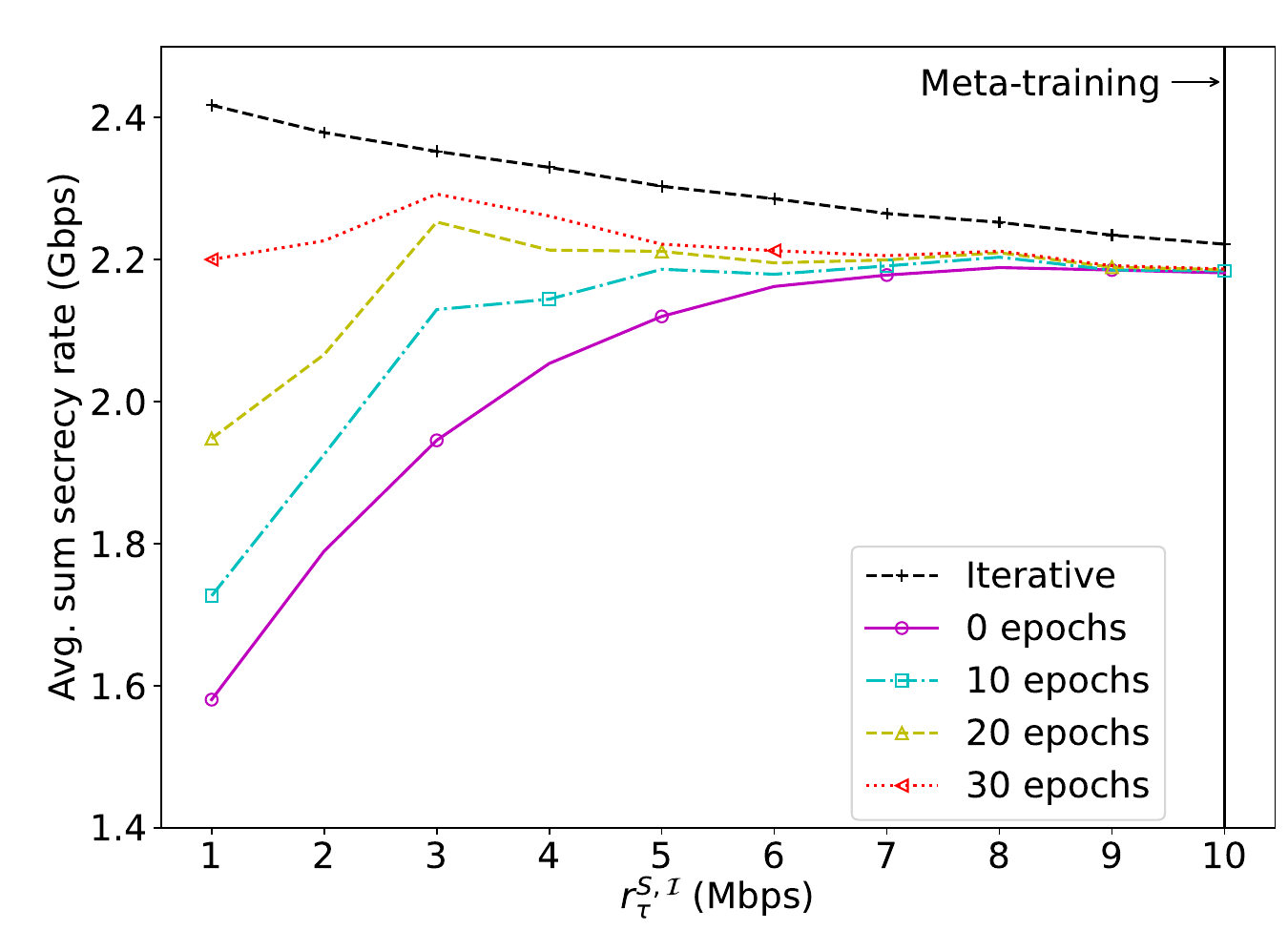}\label{fig_different_constraint_meta_10e6}}
\subfigure[$r_{\tau}^{S,\mathcal{I}}=1$~Mbps in meta-training.] 
{\includegraphics[width=9cm, height=6.3cm]{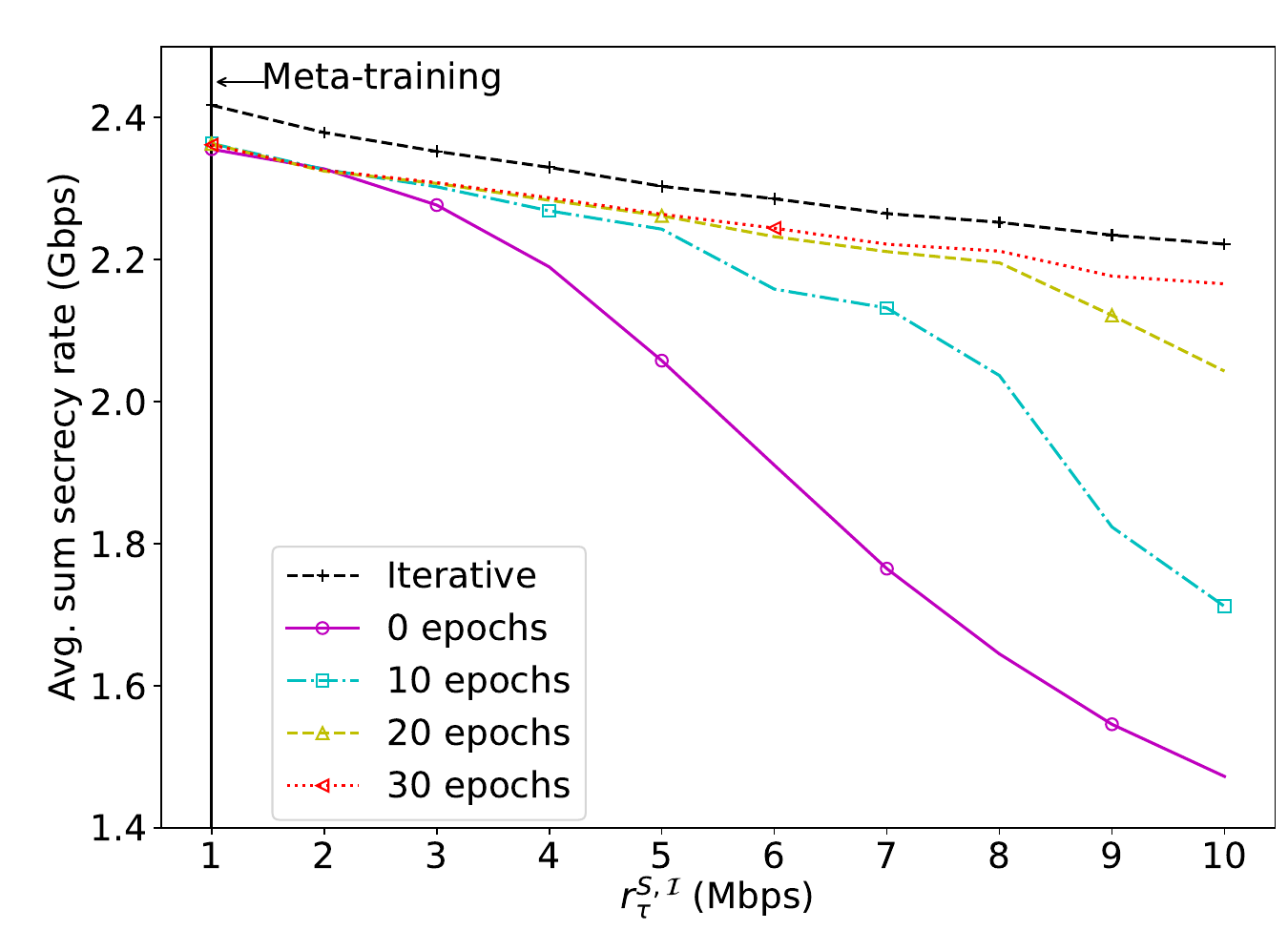}\label{fig_different_constraint_meta_1e6}}
\subfigure[$r_{\tau}^{S,\mathcal{I}} \in \{1, \cdots, 10\}$~Mbps in meta-training.]
{\includegraphics[width=9cm, height=6.3cm]{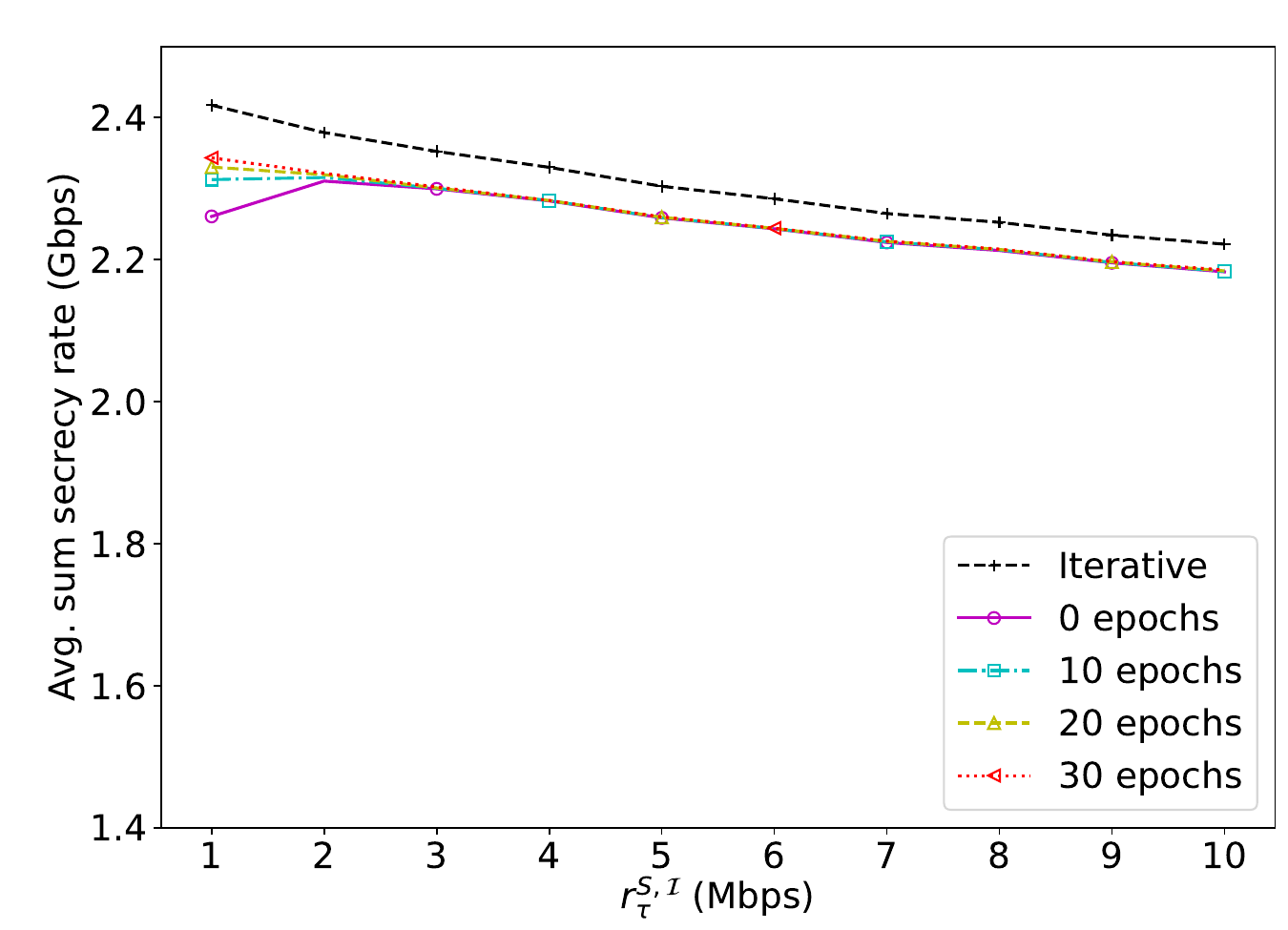}\label{fig_different_constraint_more}}
\caption{Meta-testing with dynamic secrecy rate requirements, $r_{\tau}^{S,\mathcal{I}}\in \{1, \cdots, 10\}$~Mbps, where$W_{\tau}^{S,\mathcal{I}}=100$~MHz and $U_{\tau}^{S,\mathcal{I}}=10$.}
\label{fig_mismatch_Rs_min}
\end{figure}

\subsubsection{Meta-Testing with Different System Parameters} In this part, we focus on secrecy rates in the long blocklength regime in both meta-training and meta-testing, and change the values of $r_{\tau}^{S,\mathcal{I}}$, $W_{\tau}^{S,\mathcal{I}}$, and $U_{\tau}^{S,\mathcal{I}}$ to investigate their impacts on the initial performance and sample efficiency of HML in meta-testing.
\begin{figure}[t]
\centering 
%
%
%
{\includegraphics[width=9cm, height=6.3cm]{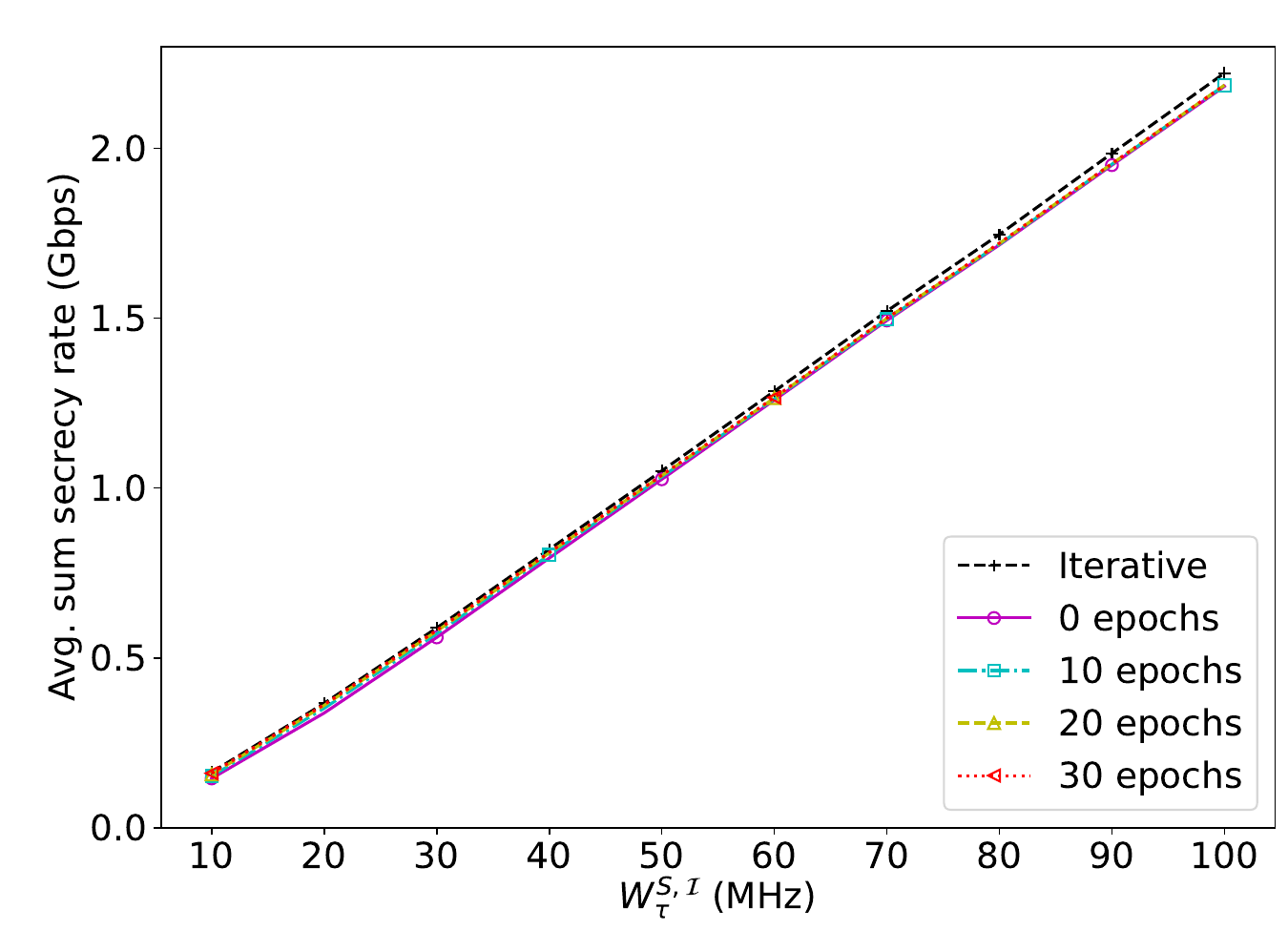}\label{fig_different_W_max_more}}
\caption{Meta-testing with dynamic bandwidth $W_{\tau}^{S,\mathcal{I}} \in \{10, \cdots, 100\}$~MHz in meta-training, where $r_{\tau}^{S,\mathcal{I}}=10$~Mbps and $U_{\tau}^{S,\mathcal{I}}=10$.}
\label{fig_sim_mismatch_W_max}
\end{figure}

\begin{figure}[t]
    \centering
    \includegraphics[width=9cm, height=6.3cm]{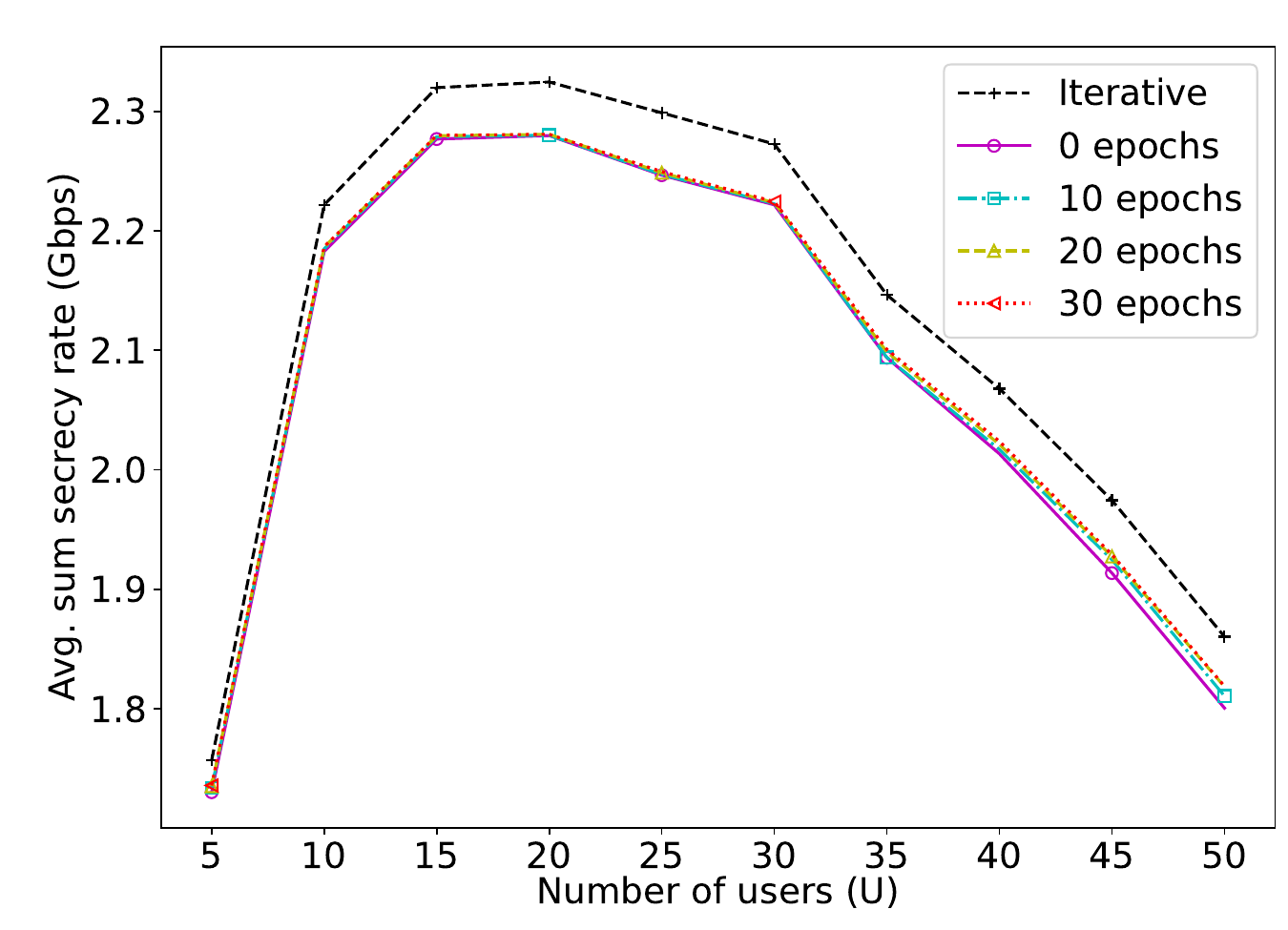}
    \caption{Meta-testing with different numbers of users $U_{\tau}^{S,\mathcal{I}} \in \{5,10, \cdots, 50\}$, where $r_{\tau}^{S,\mathcal{I}}=10$~Mbps and $W_{\tau}^{S,\mathcal{I}}=100$~MHz.}
    \label{fig_sim_mismatch_U}
\end{figure}

In Fig.~\ref{fig_mismatch_Rs_min}, we evaluate the initial performance and sample efficiency with different $r_{\tau}^{S,\mathcal{I}}$ in support sets and query sets in meta-training. Specifically, we set $r_{\tau}^{S,\mathcal{I}}$ to $10$~Mbps and $1$~Mbps in meta-training in Figs.~\ref{fig_different_constraint_meta_10e6} and~\ref{fig_different_constraint_meta_1e6}, respectively. In Fig. \ref{fig_different_constraint_more}, $r_{\tau}^{S,\mathcal{I}}$ is randomly selected from the set $\{1,\cdots,10\}$~Mbps in meta-training. In meta-testing, we increase $r_{\tau}^{S,\mathcal{I}}$ from $1$~Mbps to $10$~Mbps. The results in Figs.~\ref{fig_different_constraint_meta_10e6} and~\ref{fig_different_constraint_meta_1e6} indicate that the gaps between zero-shot learning (with $0$ training epochs in meta-testing) and the optimal solution increase with the difference between $r_{\tau}^{S,\mathcal{I}}$ in meta-training and $r_{\tau}^{S,\mathcal{I}}$ in meta-testing. To increase the generalization ability, we can increase the diversity of tasks in meta-training as shown in Fig.~\ref{fig_different_constraint_more}. In this way, our GNN is near-optimal with zero-shot learning.


In Fig.~\ref{fig_sim_mismatch_W_max}, we validated the generalization ability of our {HML} with dynamic bandwidth $W_{\tau}^{S,\mathcal{I}}$. In meta-training, $W_{\tau}^{S,\mathcal{I}}$ is randomly selecting from the set $\{10,\cdots, 100\}$~MHz. In meta-testing, we increase $W_{\tau}^{S,\mathcal{I}}$ from $10$ to $100$~MHz. The results in Fig.~\ref{fig_sim_mismatch_W_max} show that our GNN is near-optimal with different values of $W_{\tau}^{S,\mathcal{I}}$.

In Fig.~\ref{fig_sim_mismatch_U}, we further validate the generalization ability of our GNN with different numbers of users. 
In meta-training, the number of total users is randomly selected, $U_{\tau}^{S,\mathcal{I}} \in \{10,11,...,30\}$. In meta-testing, we increase the number of total users from $5$ to $50$. The results in Fig.~\ref{fig_sim_mismatch_U} show that the proposed HML can obtain a GNN that has strong generalization ability with different numbers of users, {which is consistent with our analysis in Section~\ref{section_GNN}, indicating that parameters of the user vertices are shared and reused, and the trained GNN processes the aggregated information of the scheduled users. This figure also validates that the trained GNN can effectively be applied in networks with user numbers either smaller or larger than the number of users in the meta-training set.} The gap between the GNN and the optimal policy increases slightly with $U_{\tau}^{S,\mathcal{I}}$. This is because the scale of the problem increases with $U_{\tau}^{S,\mathcal{I}}$, and it is more difficult to learn the bandwidth allocation policy of a large-scale problem compared with that of a small-scale problem.

\begin{figure}[t]
    \centering
    \includegraphics[width=9cm, height=6.3cm]{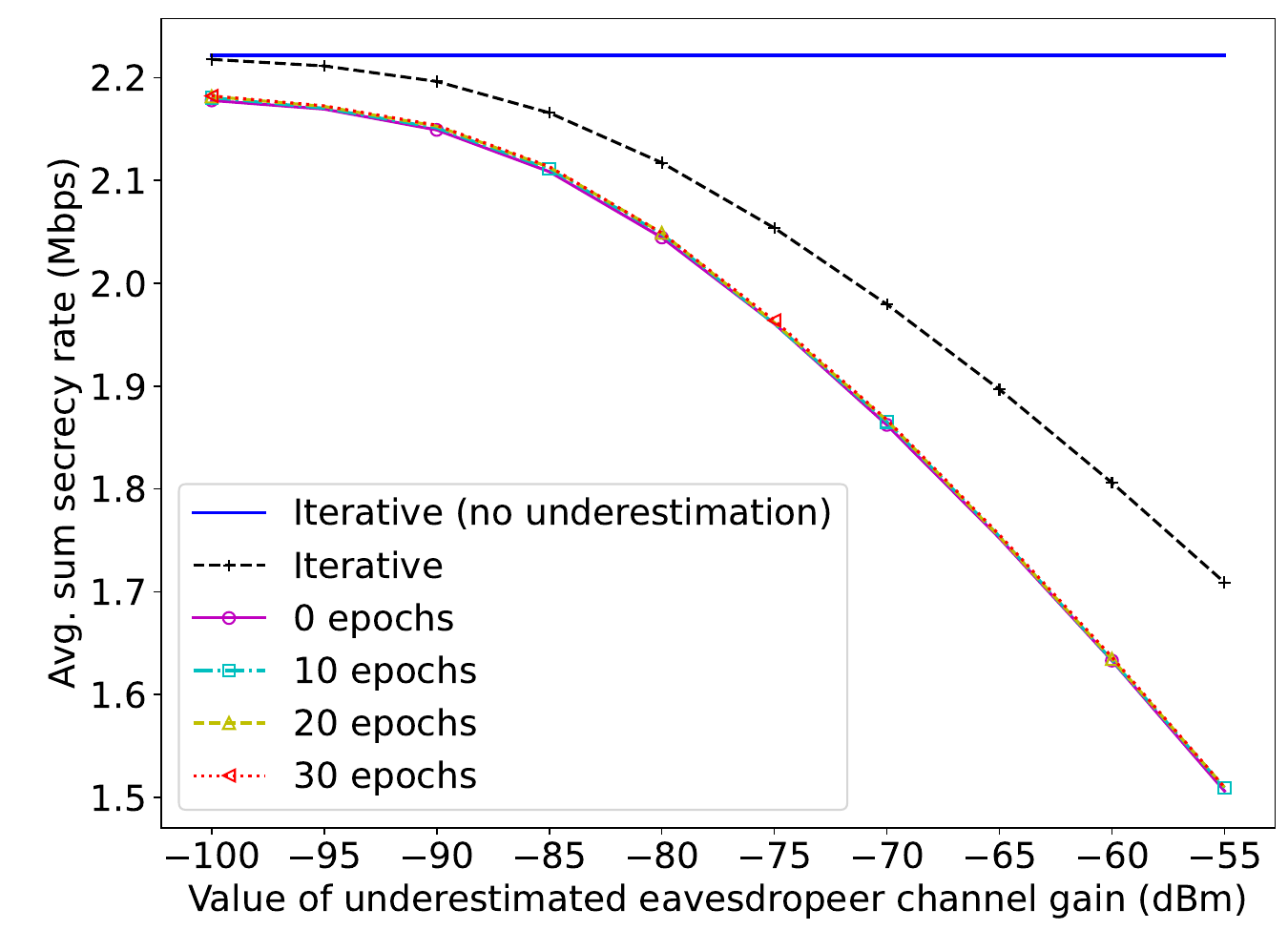}
    \caption{{Meta-testing with underestimated eavesdropper channel gain, where $r_{\tau}^{S,\mathcal{I}}=10$~Mbps, $W_{\tau}^{S,\mathcal{I}}=100$~MHz, and $U_{\tau}^{S,\mathcal{I}}=10$.}}
    \label{fig_sim_eve_uncert}
\end{figure}
{In Fig.~\ref{fig_sim_eve_uncert}, we present simulation results assessing the robustness of our HML algorithm when eavesdropper channel gain is underestimated. This figure reveals a decrease in secrecy rate as the degree of underestimation in the eavesdropper channel gain increases, aligning with the behavior predicted by eq.~\eqref{eq_secrecy_rate}.
We can also observe from this figure that the performance gap between the optimal iterative algorithm and our HML increases with the value of underestimation of the eavesdropper channel gain. Thus, there is a trade-off between the eavesdropper channel's accuracy and the choosing of the bandwidth allocation algorithms. Specifically, the optimality gap of our HML algorithm is small when the eavesdropper channel uncertainty is small, so our HML algorithm could save computation time with a tolerable performance loss. However, this optimality gap increases with the eavesdropper channel uncertainty, so we need to carefully evaluate the accuracy and the computation time when choosing the bandwidth allocation algorithm. This trade-off deserves further examination in the design and deployment of secure communication systems, particularly in real-world scenarios where perfect knowledge of the eavesdropper channel may be unattainable and where bandwidth allocation requires high computational complexity.
}

\begin{table}[t] 
\renewcommand\arraystretch{1}
\caption{{Average Computation Time (millisecond)}}
\vspace{-7 pt}
\centering 
{
\begin{tabular}{l | m{2.43cm}<{\centering} | m{1.9cm}<{\centering} | m{1.6cm}<{\centering}}
\toprule 
\toprule 
\textbf{Scenario}                               &\textbf{HML meta-training (one epoch)} &\textbf{HML inference (one sample)} &\textbf{ Iterative $\ $ (one sample)}{\centering}\\
\midrule
Fig.~\ref{fig_maximize_L_sum_secrecy_rate}      &72.61                      &3.11                   &641.62\\
\hline
Fig.~\ref{fig_maximize_S_sum_secrecy_rate}      &73.03                      &3.12                   &991.47\\
\hline
Fig.~\ref{fig_maximize_L_sum_data_rate}         &103.81                     &3.25                   &659.54\\
\hline
Fig.~\ref{fig_maximize_S_sum_data_rate}         &92.16                      &3.52                   &2325.60\\
\hline
Fig.~\ref{fig_maximize_L_sum_effective_rate}    &78.28                      &3.62                   &5392.97\\
\hline
Fig.~\ref{fig_maximize_S_sum_effective_rate}    &74.10                      &3.11                   &8266.69\\
\bottomrule 
\bottomrule 
\end{tabular}
}
\label{table_execution_time} 
\end{table}
\subsection{{Numerical Results of Computation Time}}
{Table~\ref{table_execution_time} shows the natural time of program execution, and the Python script is executed on a MacBook Pro with 1.4GHz quad-core intel core i5
and 16GB 2133 MHz LPDDR3. For the HML meta-training, we execute $2,000$ training epochs and provide the average computation time of one epoch. For the HML inference and iterative algorithm, we calculate ``sample 1'' shown in Fig.~\ref{fig_satisfy_Rs_min} for $2000$ times, and provide the average computation time of this sample. We can observe that the time required for HML inference is significantly less than the HML meta-training. This is because meta-training involves extensive optimization across large datasets, adjusting parameters to learn patterns by backpropagation, but inference simply applies the learned parameters to make predictions on new data, which requires much lighter computations. We can also observe from this table that our HML reduces the computation time by approximately $200$ to $2000$ than the optimal iterative algorithm. 
%
%
Interestingly, in the iterative algorithm, the computation times for secrecy rates are consistently shorter than those for data rates. This is primarily due to the fewer number of users scheduled in the secrecy rates-related scenarios depicted in Fig.~\ref{fig_sim_objective_secrecy_rate} compared to the data rates-related scenarios shown in Fig.~\ref{fig_sim_objective_data_rate}, leading fewer iterations required for calculating the secrecy rates.}

\section{Conclusion}
In this paper, we developed an HML approach to train a GNN-based scalable bandwidth allocation policy that can generalize well in various communication scenarios, including different number of users, wireless channels, QoS requirements, and bandwidth. The main idea is to train the initial parameters of the GNN with various tasks in meta-training, and then fine-tune the parameters with a few samples in meta-testing. Simulation results showed that the performance gap between the GNN and the optimal policy obtained by an iterative algorithm is less than $5$\% in most of the cases. For unseen communication scenarios, the GNN can converge in $10$ to $30$ training epochs, which are much faster than the existing benchmarks. {Numerical results validate that our HML can reduce the computation time by $200$ to $2000$ times compared with the optimal iterative algorithm.} Our approach can be extended beyond bandwidth allocation, such as power allocation, precoding, and repetitions. Nevertheless, the feature engineering and the structure of GNN in other scenarios deserve further investigation.

\appendices





\renewcommand{\theequation}{A.\arabic{equation}}
\setcounter{equation}{0}
\section{Proof of Concavity for Secrecy Rate in Long Blocklength Regimes~\label{appendix_concave}}
To prove the concavity of the secrecy rate in long blocklength regimes, we only need to prove that the second derivative of the secrecy rate is positive. We first calculate the partial derivative of the secrecy rate of the $k$-th scheduled user as follows,
\begin{equation}
\begin{split}
\frac{\partial r_k^{S,\mathcal{I}}(w_{\tau,k}^{D,\mathcal{I}})}{\partial {w_{\tau,k}^{D,\mathcal{I}}}}
=&\frac{\partial  \left(r_k^{D,\mathcal{I}} (w_{\tau,k}^{D,\mathcal{I}})- r_k^{e,\mathcal{I}}(w_{\tau,k}^{D,\mathcal{I}}) \right) }{\partial w_{k}}\\
=&\frac{- \zeta_k +\ln\left(1+\frac{\zeta_k^{} }{w_{\tau,k}^{D,\mathcal{I}}}\right)(w_{\tau,k}^{D,\mathcal{I}} +\zeta_k^{} )}{\ln(2)(w_{\tau,k}^{D,\mathcal{I}} + \zeta_k^{} )} \\
   &- \frac{- \zeta_k^{e} +\ln\left(1+\frac{\zeta_k^{e}}{w_{\tau,k}^{D,\mathcal{I}}}\right)(w_{\tau,k}^{D,\mathcal{I}} +\zeta_k^{e})}{\ln(2)(w_{\tau,k}^{D,\mathcal{I}} + \zeta_k^{e})} \\
=&\frac{\ln\left(\dfrac{w_{\tau,k}^{D,\mathcal{I}} + \zeta_k^{}}{w_{\tau,k}^{D,\mathcal{I}}+\zeta_k^{e}}\right) }{\ln(2)}\\
   &+\frac{ (\zeta_k^{e} - \zeta_k^{} ) w_{\tau,k}^{D,\mathcal{I}}}{\ln(2)(w_{\tau,k}^{D,\mathcal{I}}+\zeta_k^{}  )(w_{\tau,k}^{D,\mathcal{I}}+\zeta_k^{e})}
,
\end{split}
\end{equation}
\normalsize
where $\zeta_k^{} = {P_k h_k}/{N0}$ and $\zeta_k^{e} = {P_k h_k^e}/{N0}$. Since the secrecy rate of the user increases with the increasing of the allocated bandwidth, we have ${\partial r_k^{S,\mathcal{I}}(w_{\tau,k}^{D,\mathcal{I}})}/{\partial {w_{\tau,k}^{D,\mathcal{I}}}}<0$. 
%
The second derivative of $r_k^{S,\mathcal{I}}(w_{\tau,k}^{D,\mathcal{I}})$ can be derived as follows,
\begin{equation}
\begin{split}
\frac{\partial^{2} r_k^{S,\mathcal{I}}(w_{\tau,k}^{D,\mathcal{I}})}{\partial {w_{\tau,k}^{D,\mathcal{I}}}^2}
=&\frac{\partial  }{\partial w_{\tau,k}^{D,\mathcal{I}}} \left(\frac{\partial r_k^{S,\mathcal{I}}(w_{\tau,k}^{D,\mathcal{I}})}{\partial {w_{\tau,k}^{D,\mathcal{I}}}}\right)\\
=&\dfrac{(\zeta_k^{e} - \zeta_k)\left((\zeta_k^{e} + \zeta_k)w_{\tau,k}^{D,\mathcal{I}} + 2\zeta_k^{e} \zeta_k\right)}
{\ln(2) (w_{\tau,k}^{D,\mathcal{I}} + \zeta_k)^2 (w_{\tau,k}^{D,\mathcal{I}} + \zeta_k^{e})^2}
.
\end{split}
\end{equation}
\normalsize
For any scheduled user, we have $\zeta_k  > \zeta_k^{e}$. Thus, ${\partial^{2} r_k^{S,\mathcal{I}}(w_{\tau,k}^{D,\mathcal{I}})}/{\partial {w_{\tau,k}^{D,\mathcal{I}}}^2} < 0$. Therefore, $r_k^{S,\mathcal{I}}(w_{\tau,k}^{D,\mathcal{I}})$ is concave. This completes the proof. {\hfill $\square$\par}
%
\footnotesize

\begin{IEEEbiography}
[{\includegraphics[width=1in,height=1.25in,clip,keepaspectratio]{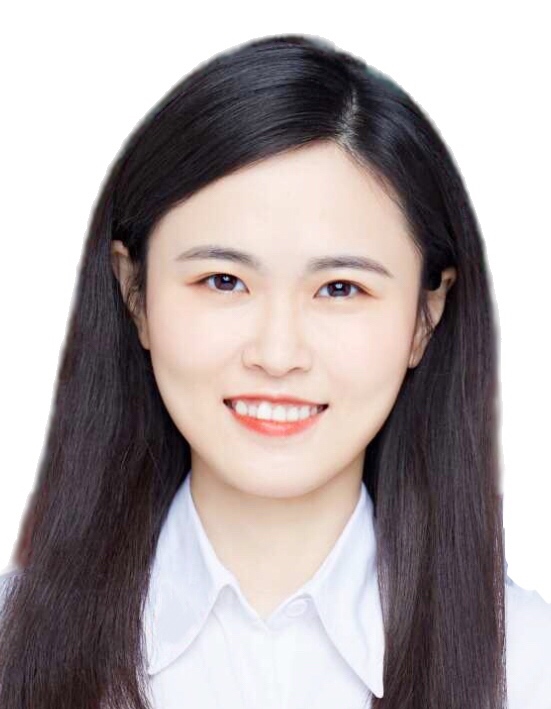}}] 
{Xin Hao} received her B.S. and M.E. degrees from the University of Electronic Science and Technology of China (UESTC) in 2012 and 2015, respectively, and her Ph.D. degree from The University of Sydney (USYD) in 2024. She is currently a Research Fellow with the School of Information Technology in Artificial Intelligence at Deakin University. She is a recipient of the USYD 2023 Faculty of Engineering PhD Completion Award, the 2023 Faculty of Engineering Research Scholarship, and the 2020 Faculty of Engineering Research Scholarship. Her research interests include multi-task learning, multi-agent reinforcement learning, meta-learning, and security in low-latency Internet-of-Things networks. She served as a session chair in the 2023 IEEE International Conference on Communications (ICC) workshop and a technical committee member of the 2024 IEEE International Conference on Robotics and Automation (ICRA) workshop.
\end{IEEEbiography} 

\begin{IEEEbiography}[{\includegraphics[width=1in,height=1.25in,clip,keepaspectratio]{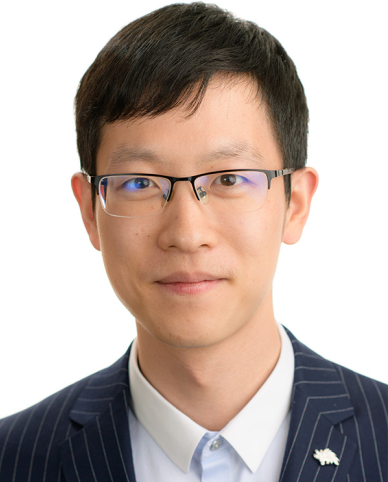}}] 
{Changyang~She} (S'12–M'17–SM'23) received his B. Eng degree in Honors College of Beihang University, Beijing, China in 2012 and Ph.D. degree from the School of Electronics and Information Engineering of Beihang University in 2017. From 2017 to 2018, he was a postdoctoral research fellow at the Singapore University of Technology and Design. From 2018 to 2021, he was a postdoctoral research associate at the University of Sydney. From 2021 to 2024, he served as the Australian Research Council (ARC) Discovery Early Career Research Award (DECRA) fellow (Level B) at the University of Sydney. He is currently a full professor at Harbin Institute of Technology (Shenzhen). His research interests lie in the areas of ultra-reliable and low-latency communications (URLLC), deep learning in wireless networks, edge intelligence, energy-efficient communications, and interdisciplinary research in the metaverse. 
\end{IEEEbiography} 

\begin{IEEEbiography}[{\includegraphics[width=1in,height=1.25in,clip,keepaspectratio]{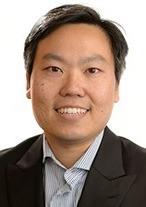}}] 
{Phee~Lep~Yeoh} (S'08-M'12–SM'23) received the B.E. degree with University Medal and the Ph.D. degree from the University of Sydney, Australia, in 2004 and 2012, respectively. From 2012 to 2016, he was a Lecturer at the University of Melbourne, Australia, and from 2016 to 2023, he was a Senior Lecturer at the University of Sydney, Australia. In 2023, he joined the School of Science, Technology, and Engineering at University of the Sunshine Coast in Queensland, Australia. Dr Yeoh is a recipient of the 2020 University of Sydney Robinson Fellowship, the 2018 Alexander von Humboldt Research Fellowship for Experienced Researchers, and the 2014 Australian Research Council Discovery Early Career Researcher Award. He has also received best paper awards at IEEE PIMRC 2023, IEEE ICC 2014 and IEEE VTC Spring 2013.
\end{IEEEbiography} 

\begin{IEEEbiography}[{\includegraphics[width=1in,height=1.25in,clip,keepaspectratio]{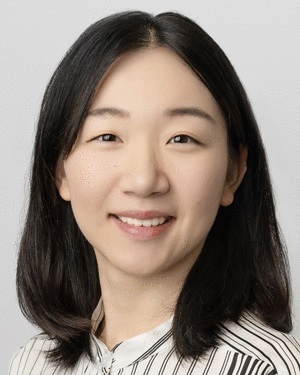}}] 
{Yuhong~Liu} received the B.Eng. degree in telecommunications engineering with management from the Beijing University of Posts and Telecommunication, Beijing, China, in 2014, and the Ph.D. degree in electrical engineering from the University of Sydney, Sydney, NSW, Australia, in 2022. Her research interests include ultra-reliable and low-latency communications, deep learning and meta-learning in wireless networks, and graph neural networks. She was the recipient of the University of Sydney Postgraduate Award (UPA).
\end{IEEEbiography} 

\begin{IEEEbiography}[{\includegraphics[width=1in,height=1.25in,clip,keepaspectratio]{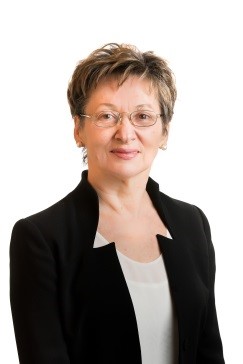}}]
{Branka Vucetic} (Life Fellow, IEEE) is currently an ARC Laureate Fellow and Director of the Centre of Excellence for IoT and Telecommunications at the University of Sydney. Her current work is in the areas of wireless networks and Internet of Things. In the area of wireless networks, she works on communication system design for millimetre wave (mmWave) frequency bands. In the area of the Internet of things, she works on providing wireless connectivity for mission critical applications. She is a Fellow of the Australian Academy of Science, the Australian Academy of Technological Sciences and Engineering, and the Engineers Australia.
\end{IEEEbiography}

\begin{IEEEbiography}[{\includegraphics[width=1in,height=1.25in,clip,keepaspectratio]{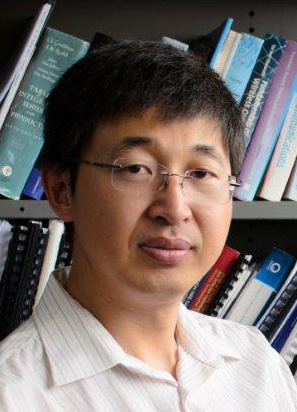}}]
{Yonghui~Li} (Fellow, IEEE) received the Ph.D.
degree from Beijing University of Aeronautics and Astronautics, in November 2002. Since 2003, he has been with the Centre of Excellence in Telecommunications, The University of Sydney, Australia. He is now a Professor and Director of Wireless Engineering Laboratory in School of Electrical and Computer Engineering, The University of Sydney. He was the recipient of the Australian Queen Elizabeth II Fellowship in 2008 and the Australian Future Fellowship in 2012. His current research interests include wireless communications, with a particular focus on MIMO, millimeter wave communications, machine to machine communications, coding techniques, and cooperative communications. He holds a number of patents granted and pending in these fields. He was an Editor for IEEE TRANSACTIONS ON COMMUNICATIONS and IEEE TRANSACTIONS ON VEHICULAR TECHNOLOGY. He also served as the Guest Editor for several IEEE journals, such as IEEE JOURNAL ON SELECTED AREAS IN COMMUNICATIONS, IEEE COMMUNICATIONS MAGAZINE, IEEE INTERNET OF THINGS JOURNAL, and IEEE ACCESS. He received the best paper awards from IEEE International Conference on Communications (ICC) 2014, IEEE PIMRC 2017, and IEEE Wireless Days Conferences (WD) 2014.
\end{IEEEbiography}

\end{document}